\documentclass[paper]{ptephy_v1}

\usepackage{boites}
\usepackage{graphicx}
\usepackage{bm}
\usepackage{amsmath,amssymb,amsfonts}

\newcommand{\nc}{\newcommand}		
\nc{\vc}[1]	{\mbox{\boldmath $#1$}}	
\nc{\del}       {\partial}              
\nc{\bra}       {\langle}               
\nc{\ket}       {\rangle}               
\nc{\beq}       {\begin{eqnarray}}
\nc{\eeq}       {\end{eqnarray}}
\nc{\AMD}       {{\rm AMD}}
\nc{\TOAMD}     {{\rm TOAMD}}

\begin{document}

\title{New successive variational method of tensor-optimized antisymmetrized molecular dynamics for nuclear many-body systems}

\author{\name{Takayuki Myo}{1,2}, \name{Hiroshi Toki}{2},  \name{Kiyomi Ikeda}{3}, \name{Hisashi Horiuchi}{2}, and \name{Tadahiro Suhara}{4}}

\address{\affil{1}{General Education, Faculty of Engineering, Osaka Institute of Technology, Osaka, Osaka 535-8585, Japan}
\affil{2}{Research Center for Nuclear Physics (RCNP), Osaka University, Ibaraki, Osaka 567-0047, Japan}
\affil{3}{RIKEN Nishina Center, Wako, Saitama 351-0198, Japan}
\affil{4}{Matsue College of Technology, Matsue 690-8518, Japan}
\email{takayuki.myo@oit.ac.jp}}

\begin{abstract}%
We recently proposed a new variational theory of ``tensor-optimized antisymmetrized molecular dynamics'' (TOAMD),
which treats the strong interaction explicitly for finite nuclei [T. Myo et al., Prog. Theor. Exp. Phys. {\bf 2015}, 073D02 (2015)] .
In TOAMD, the correlation functions for the tensor force and the short-range repulsion and their multiple products are successively operated to the AMD state.
The correlated Hamiltonian is expanded into many-body operators by using the cluster expansion and 
all the resulting operators are taken into account in the calculation without any truncation.
We show detailed results for TOAMD with the nucleon--nucleon interaction AV8$^\prime$ for $s$-shell nuclei. 
The binding energy and the Hamiltonian components are successively converged to exact values of the few-body calculations.
We also apply TOAMD to the Malfliet--Tjon central potential having a strong short-range repulsion.
TOAMD can treat the short-range correlation and provided accurate energies of $s$-shell nuclei, reproducing the results of few-body calculations.
It turns out that the numerical accuracy of TOAMD with double products of the correlation functions is beyond the variational Monte Carlo method with Jastrow's product-type correlation functions.
\end{abstract}

\subjectindex{D10, D11}


\maketitle

\section{Introduction}

The nucleon--nucleon ($NN$) interaction has two characteristics: 
strong repulsion at short distances and a strong tensor force at long and intermediate distances \cite{pieper01,wiringa95}.
The latter dominantly comes from the one-pion exchange between nucleons.
These characteristics of the $NN$ interaction provide the high-momentum components of nuclear motion in nuclei, which should be included in the nuclear wave function. 
The short-range repulsion reduces the amplitudes of a short-distance nucleon pair in a nucleus.
The tensor force causes the $D$-wave state of a nucleon pair, induced by the strong $S$--$D$ coupling of the tensor force.
This $D$-wave state has a spatially compact property as compared with the $S$-wave state owing to the high-momentum component of the tensor correlation \cite{ikeda10,ong13}. 

We have treated the tensor and short-range correlations in nuclei with two types of theoretical methods.
One is the shell-model-type approach, which we call the ``tensor-optimized shell model'' (TOSM) \cite{myo05,myo07,myo09}.
In TOSM, we optimize the two-particle--two-hole (2p--2h) states fully in the wave function without any truncation for the particle states.
These 2p--2h excitations can describe the strong tensor correlation in nuclei via the coupling between 0p--0h and 2p--2h configurations.
For the short-range correlation, it is in general difficult to express this correlation in the shell model based on the single-particle picture.
We have combined TOSM with the central part of the unitary correlation operator method (UCOM) in which the central-type short-range correlation is explicitly treated \cite{feldmeier98,myo09}. 
The central-type shift operator is introduced in order to reduce the short-range amplitudes of the relative motion of a nucleon pair in nuclei.
In UCOM, the transformed Hamiltonian is truncated up to two-body operators, while the exact transformation leads to many-body operators.
The two-body approximation of UCOM is considered to be reasonable for the short-range correlation.
In TOSM+UCOM, we explicitly treat the tensor and short-range correlations in the wave function.
This method nicely describes the shell-model-like states with the correct order of energy levels in the $p$-shell nuclei \cite{myo11,myo12,myo14,myo15b}.

On the other hand, nuclear clustering is an important aspect of nuclear structure, such as the triple-$\alpha$ Hoyle state in $^{12}$C \cite{ikeda68,horiuchi12}.
The clustering correlation is difficult to treat in the shell-model-type approach \cite{barrett13,myo14}.
Recently, we have developed a new variational theory for the clustering description of nuclei using the $NN$ interaction \cite{myo15,myo16,myo17}.
We employ antisymmetrized molecular dynamics (AMD) \cite{kanada03,kanada12} as the basis state to describe the clustering states,
and introduce two correlation functions of tensor-operator type and central-operator type to treat the $NN$ interaction. 
The correlation functions are multiplied by the AMD wave function and superposed with the original AMD wave function.
We call this method ``tensor-optimized antisymmetrized molecular dynamics'' (TOAMD) \cite{myo15}.
The concept of TOAMD is similar to TOSM in treating the tensor and short-range correlations induced by the $NN$ interaction.
The scheme of TOAMD is extendable increasing the series of the multiple products of the two correlation functions by successive power expansion.

In TOAMD, there appear multiple products of the Hamiltonian and the correlation functions, which become the series of many-body operators in the cluster expansion.
We treat all the resulting many-body operators, which makes TOAMD a variational method.
In the previous work \cite{myo16}, we have described the $s$-shell nuclei with TOAMD within the double products of the correlation functions,
and have shown that TOAMD nicely reproduces the results of the Green's function Monte Carlo (GFMC) using the AV8$^\prime$ bare $NN$ interaction \cite{kamada01}.  
The TOAMD provides a general formulation for nuclei with various mass numbers.
The concept of TOAMD, which introduces the products of correlation functions successively, can be applied to the nuclear matter problem.
Recently Toki and Hu have formulated tensor-optimized relativistic nuclear matter \cite{toki17}.
They proposed a variational framework for nuclear matter starting from the $NN$ interaction, in which
the correlation functions are multiplied successively by the plane-wave basis states. This framework is conceptually the same as TOAMD except for the basis functions.

We have also applied TOAMD to the description of the central-type correlation coming from the central $NN$ interaction 
which has a short-range repulsion \cite{myo17}, in the same scheme as the previous work \cite{myo16}.
This subject gives scope to the application of TOAMD to other fields.
We have focused on the central-type correlation and use the Malfliet--Tjon V (MT-V) central potential with a Yukawa-type tail and a strong short-range repulsion \cite{malfliet69}.
We have shown that the results of TOAMD for $s$-shell nuclei are good as compared with the few-body calculation.
We have also compared the results of TOAMD with those using UCOM, which showed that UCOM well describes the short-range correlations of nuclei quantitatively under the two-body approximation of the unitary transformation.

In this paper, we perform a detailed analysis of $s$-shell nuclei with the AV8$^\prime$ bare $NN$ interaction 
and the MT-V central interaction in TOAMD, the basic results of which are reported in Refs.~\cite{myo16,myo17}.
We explain the detailed procedure of the TOAMD calculation for finite nuclei.
We give the classification of the many-body operators in the cluster expansion of the correlated operators by using a diagrammatic representation.
We also explain how to perform the energy variation including the multiple products of the correlation functions using Gaussian expansion in the TOAMD wave function.
We report the properties of the TOAMD wave function, such as the spatial distribution of the correlation functions and the role of the many-body operators arising from the correlated Hamiltonian.
The individual contributions of the many-body operators are discussed for each Hamiltonian component.
In TOAMD, the wave function has a form of power series expansion with respect to the correlation functions, where each correlation function is determined independently in the energy minimization.
This property of TOAMD is different from the ordinary variational approach with the Jastrow ansatz, in which the common correlation function is multiplied by every pair in nuclei.
We discuss the detailed effect of the successive and independent optimization of the correlation functions on the solutions of TOAMD.

In Sect.~\ref{sec:method}, we explain the essential features of TOAMD.  In particularly, we give the details of the cluster expansion leading to many-body operators.
In Sect.~\ref{sec:results}, we present the results for $s$-shell nuclei.  We show the results using the AV8$^\prime$ bare interaction and the Malfliet--Tjon central interaction.
A summary is given in Sect.~\ref{sec:summary}.

\section{Tensor-optimized antisymmetrized molecular dynamics (TOAMD)}\label{sec:method}

\subsection{Definition}
We explain here the essential features of TOAMD. The details of TOAMD are given in Ref. \cite{myo15}. We employ the AMD wave function as a reference state for TOAMD. 
The AMD wave function $\Phi_{\rm AMD}$ is given as the Slater determinant consisting of Gaussian wave packets of nucleons with mass number $A$ as follows:
\begin{eqnarray}
\Phi_{\rm AMD}
&=& \frac{1}{\sqrt{A!}} {\rm det} \left\{ \prod_{i=1}^A \phi_i \right\}~,
\label{eq:AMD}
\\
\phi(\vec r)&=&\left(\frac{2\nu}{\pi}\right)^{3/4} e^{-\nu(\vec r-\vec D)^2} \chi_{\sigma} \chi_{\tau}.
\label{eq:Gauss}
\end{eqnarray}
The single-nucleon wave function $\phi(\vec r)$ has a Gaussian wave packet with a range parameter $\nu$ and a centroid position $\vec D$, 
a spin part $\chi_{\sigma}$ and an isospin part $\chi_{\tau}$.
In the present study of $s$-shell nuclei, $\chi_{\sigma}$ is fixed with the up or down component and $\chi_{\tau}$ is the proton or neutron.
The range parameter $\nu$ is common for all nucleons. In this condition we can factorize the center-of-mass (c.m.) wave function from $\Phi_{\rm AMD}$.
In the shell-model limit with the condition of $\vec D_i=0$ ($i=1,\ldots, A$) for all nucleons,
the range $\nu$ has relation to $\hbar \omega$ as $\hbar^2 \nu/ m = \hbar\omega/2$ where $m$ is the nucleon mass.

In TOAMD, we consider two kinds of correlations induced by the tensor force and short-range repulsion, both of which are difficult to express in $\Phi_{\rm AMD}$.
Following the concept in Refs. \cite{sugie57,nagata59}, we introduce the pair-type correlation functions, $F_D$ for the tensor force and $F_S$ for the short-range repulsion. 
We multiply them by the AMD wave function individually and superpose these components with the AMD wave function. The correlation functions $F_D$ and $F_S$ are determined variationally.
This concept of TOAMD is similar to TOSM \cite{myo05,myo07,myo09}. Here we define the two-body correlation functions as
\begin{eqnarray}
F_D
&=& \sum_{t=0}^1\sum_{i<j}^A f^{t}_{D}(r_{ij})\, r_{ij}^2\, S_{12}(\hat{r}_{ij})\, (\vec \tau_i\cdot \vec \tau_j)^t \,,
\label{eq:Fd}
\\
F_S
&=& \sum_{t=0}^1\sum_{s=0}^1\sum_{i<j}^A f^{t,s}_{S}(r_{ij})\,(\vec \tau_i\cdot \vec \tau_j)^t\, (\vec\sigma_i\cdot \vec\sigma_j)^s\,,
\label{eq:Fs}
\end{eqnarray}
with a relative coordinate $\vec r_{ij}=\vec r_i - \vec r_j$.
The pair functions $f^{t}_{D}(r)$ and $f^{t,s}_{S}(r)$ are the variational parameters and are explained later. 
The labels $t$ and $s$ stand for the isospin and spin channels of the two nucleons, respectively.
The functions $F_D$ and $F_S$ change the relative motion of a nucleon pair in $\Phi_{\rm AMD}$ and do not excite the c.m. motion of $\Phi_{\rm AMD}$.  
The function $F_D$ produces the $D$-wave transition owing to the tensor operator $S_{12}$ given as
\begin{eqnarray}
S_{12}(\hat r_{ij})=3(\vec \sigma_i\cdot \hat r_{ij})(\vec \sigma_j \cdot \hat r_{ij})-\vec \sigma_i \cdot \vec \sigma_j~.
\end{eqnarray}
The functions $F_D$ and $F_S$ are scalar operators and do not change the angular-momentum state of $\Phi_{\rm AMD}$.
In general, two functions $F_D$ and $F_S$ are not commutable.
Physically, the functions $F_D$ and $F_S$ can excite two nucleons in nuclei to the high-momentum state, which corresponds to the 2p--2h excitations in TOSM \cite{myo05}.

We multiply these correlation functions by the AMD wave function.
In the present study of TOAMD, we include up to the double products of the correlation functions.
We define the TOAMD wave function with the single correlation functions as
\begin{eqnarray}
\Phi_{\rm TOAMD}^{\rm single}
&=& (1+F_S+F_D) \times\Phi_{\rm AMD}\,.
\label{eq:TOAMD1}
\end{eqnarray}
We call this form of the TOAMD wave function ``single TOAMD''.
As an extension of Eq.~(\ref{eq:TOAMD1}), we increase the order of the correlation function in TOAMD by adding the double products consisting of $F_D$ and $F_S$ following the previous study \cite{myo15}.
As with the single case, we define the TOAMD wave function with the double products of the correlation functions below, which we call ``double TOAMD'' as 
\begin{eqnarray}
\Phi_{\rm TOAMD}^{\rm double}
&=& (1+F_S+F_D+F_S F_S+F_S F_D+F_D F_S+F_D F_D) \times\Phi_{\rm AMD}~.
\label{eq:TOAMD2}
\end{eqnarray}
This form is based on the power series expansion in terms of the correlation functions $F_D$ and $F_S$.
It is noted that $F_D$ and $F_S$ in each term in Eq.\,(\ref{eq:TOAMD2}) are independent and variationally determined.
This indicates that we have five kinds of $F_D$ and five kinds of $F_S$ in the double TOAMD wave function.
For simplicity, we denote the common symbols of $F_D$ and $F_S$ in $\Phi_{\rm TOAMD}^{\rm double}$.
This wave function of TOAMD has a general form with respect to mass number $A$ and is commonly used for all nuclei and also for nuclear matter \cite{toki17}.  
As an extension of Eq.~(\ref{eq:TOAMD2}), we can successively increase the order of power expansion with $F_D$ and $F_S$ to triple products such as $F_D F_D F_S$ when we want to increase the variational accuracy of the results.

\subsection{Cluster expansion of the correlated operators}
We use the Hamiltonian with a two-body bare $NN$ interaction $V$ for mass number $A$ as
\begin{eqnarray}
    H
&=& T+V
~=~ \sum_i^{A} t_i - T_{\rm c.m.} + \sum_{i<j}^{A} v_{ij}\, ,
    \label{eq:Ham}
    \\
    v_{ij}
&=& v_{ij}^{\rm C} + v_{ij}^{\rm T} + v_{ij}^{LS}\, .
\end{eqnarray}
Here, $t_i$ and $T_{\rm c.m.}$ are the kinetic energies of each nucleon and the center-of-mass, respectively.
We employ a bare $NN$ interaction $v_{ij}$ AV8$^\prime$ \cite{pieper01} consisting of central $v^{\rm C}_{ij}$, tensor $v^{\rm T}_{ij}$, and spin--orbit $v^{LS}_{ij}$ terms, 
which is used in the benchmark calculation of $^4$He \cite{kamada01}. 
The total energy $E$ in TOAMD is given as:
\begin{eqnarray}
    E
&=& \frac{\langle\Phi_{\rm TOAMD} |H|\Phi_{\rm TOAMD}\rangle}{\langle\Phi_{\rm TOAMD} |\Phi_{\rm TOAMD}\rangle}
    \nonumber\\
&=& \frac{\langle\Phi_{\rm AMD} | H + F^\dagger H + H F + F^\dagger H F + \cdots |\Phi_{\rm AMD}\rangle}{\langle\Phi_{\rm AMD} | 1+F^\dagger +F  + F^\dagger F + \cdots |\Phi_{\rm AMD}\rangle}
    \nonumber\\
&=& \frac{\langle\Phi_{\rm AMD} | \tilde{H} |\Phi_{\rm AMD}\rangle}{\langle\Phi_{\rm AMD} | \tilde{N} |\Phi_{\rm AMD}\rangle} \, ,
\label{eq:E_TOAMD}
\end{eqnarray}
where $F$ stands for $F_D$ and $F_S$. The operators $\tilde{H}$ and $\tilde{N}$ in the last equation are the correlated Hamiltonian and norm operator, respectively.
We calculate the matrix elements of the correlated operators with the AMD wave function.
The operators $\tilde{H}$ and $\tilde{N}$ consist of the various products of correlation functions, such as $F^\dagger H F$ and $F^\dagger F$.
These operators are individually expanded into a series of many-body operators in terms of the cluster expansion,
the detailed procedure of which is given in Ref.~\cite{myo15}.
In the case of the kinetic energy $T$, $F^\dagger T F$ is expanded into many-body operators from two to five, with various combinations of particle index.
For the two-body interaction $V$, $F^\dagger V F$ is expanded into many-body operators from two- to six-body ones. 
In the same way, $F^\dagger F^\dagger V F F$ gives up to ten-body operators.

In the calculation of Eq.~(\ref{eq:E_TOAMD}), we use all the resulting many-body operators in the cluster expansion without any truncation. 
This is of importance to keep TOAMD as a variational theory.
The procedure of the calculation of the matrix elements is performed systematically for any order of the multiple products of the correlation functions and also for any mass number.
Later we explain the cluster expansion for the single TOAMD case.

In the Hamiltonian given in Eq. (\ref{eq:Ham}), the c.m. kinetic energy operator $T_{\rm c.m.}$ is not affected by the correlation function as $[T_{\rm c.m.}, F]=0$.
This property indicates that we can calculate the c.m. kinetic energy using the AMD wave function.
The AMD wave function has the $0s$ c.m. state under the condition of $\sum_{i=1}^A \vec D_i =0$ for the centroid of the c.m. position. 
Finally we obtain the matrix element of the c.m. kinetic energy using the matrix elements of the correlated norm operator as 
\begin{eqnarray}
    \bra\Phi_{\rm TOAMD} |T_{\rm c.m.}|\Phi_{\rm TOAMD} \ket
&=& \frac{3\hbar^2 \nu}{2m} \bra\Phi_{\rm TOAMD}|\Phi_{\rm TOAMD} \ket
    \nonumber\\
&=& \frac{3\hbar^2 \nu}{2m} \bra\Phi_{\rm AMD}|\tilde N|\Phi_{\rm AMD} \ket \,.
    \label{eq:CM}
\end{eqnarray}

\begin{figure}[t]
\begin{center}
\includegraphics[width=7.5cm,clip]{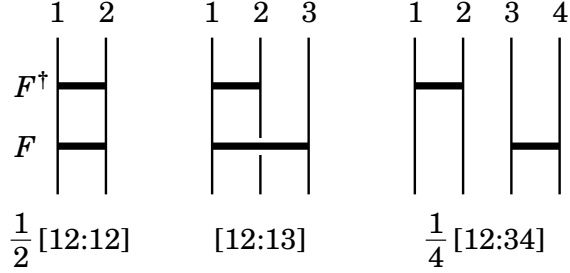}
\caption{Diagrams of the cluster expansion of $F^\dagger F$.
Vertical lines indicate the particles numbering from the left side as $1,2,3,4$. 
Short horizontal lines indicate the two-body correlation function $F$.
Each diagram represents the ladder (left), linked (middle), and unlinked (right) types.
They are two-body, three-body, and four-body operators, respectively.
We also write the configurations with symmetry factor below each diagram.}
\label{fig:FF}
\end{center}
\end{figure}

\begin{figure}[t]
\begin{center}
\includegraphics[width=13.0cm,clip]{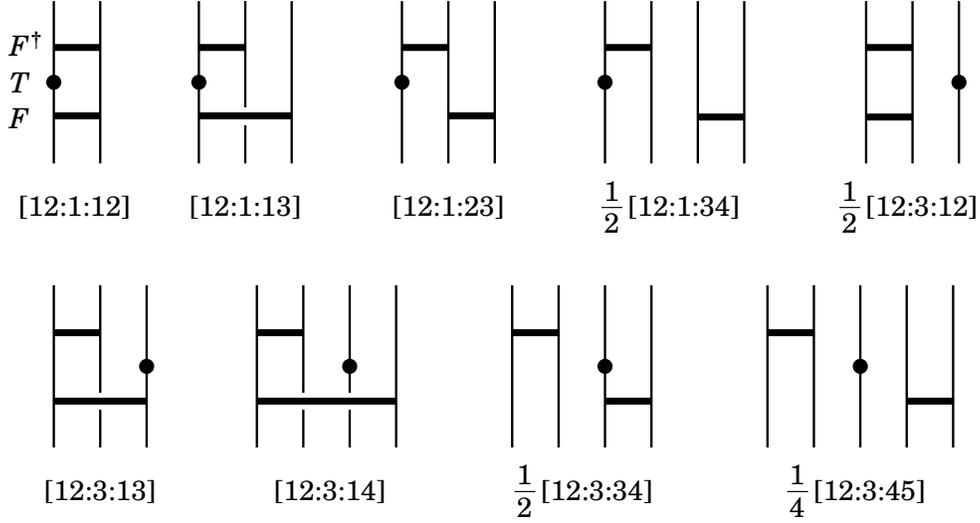}
\caption{Diagrams of the cluster expansion of $F^\dagger TF$ with the one-body kinetic energy operator $T$ indicated by solid circles.}
\label{fig:FGF}
\end{center}
\end{figure}

\begin{figure}[t]
\begin{center}
\includegraphics[width=15.0cm,clip]{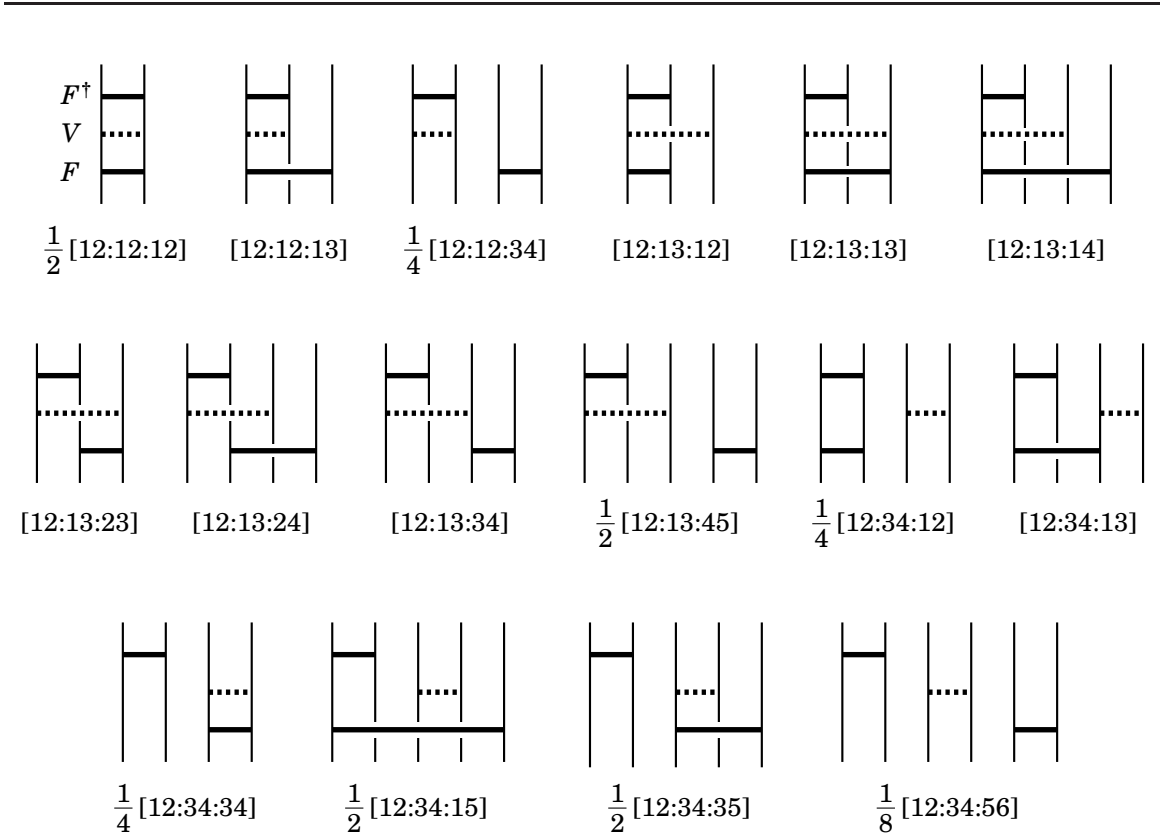}
\caption{Diagrams of the cluster expansion of $F^\dagger VF$. The dotted lines indicate the two-body interaction $V$.}
\label{fig:FFF}
\end{center}
\end{figure}

We discuss typical cases of the cluster expansion of the correlated Hamiltonian $\tilde{H}$ and norm operator $\tilde{N}$.
We show all the diagrams to be calculated in the single TOAMD in Figs.\,\ref{fig:FF}, \ref{fig:FGF}, and \ref{fig:FFF}.
We follow the rule of cluster expansion explained in Ref. \cite{myo15}, in which the two-body correlation function $F$ is symbolically expressed as
\begin{eqnarray}
  F&=&\sum_{i<j}^Af_{ij} ~=~ \frac12 \sum_{i\ne j}^Af_{ij}~\Rightarrow~\frac12 [12]\,,
  \label{eq:12}
\end{eqnarray}
where the factor $1/2$ is a symmetry factor to set the number of interaction pairs in $F$. 
Hereafter we write the condition of ``$i\ne j$'' as ``$i,j$'' in the summation of the particle index for simplicity.
The notation of the configuration with square brackets $[12]$ indicates the term of $\sum^A_{i,j} f_{ij}$, 
where the particle index $i$ and $j$ correspond to 1 and 2 in the configuration, respectively. We notice that the symmetry $[12]=[21]$ from the particle exchange.
In addition, we put the symmetry factor $1/2$ in front of the configuration to express the function $F$ in Eq.~(\ref{eq:12}).

Following the rule of this expression for $F$, we expand the double product $F^\dagger F$, which appears in the correlated norm operator $\tilde N$,
into three terms as 
\begin{eqnarray}
  F^\dagger F&=& \left( \frac12 \sum_{i,j}f^\dagger_{ij}\right) \left( \frac12 \sum_{k,l}f_{kl}\right)
  \\
&=& \frac12 \sum_{i,j}f^\dagger_{ij} f_{ij} + \sum_{i,j,k}f^\dagger_{ij}f_{ik} + \frac14 \sum_{i,j,k,l} f^\dagger_{ij}f_{lk}
  \\
&\Rightarrow&\frac12 [12:12] + [12:13] + \frac14 [12:34]
  \label{eq:FF}
\end{eqnarray}
We again put the symmetry factor in front of the configuration with the square brackets.
When the symmetry factor is unity, we do not show it.
In general, the symmetry factor has a value of $\left(1/2\right)^N$, where $N$ is the number of symmetry with respect to the particle exchange in the configuration.
When there is no symmetry, $N=0$ and the symmetry factor becomes unity.

In Eq.~(\ref{eq:FF}), the colon in the configurations represents the partition between the different correlation functions. 
In this case, we obtain three kinds of configurations: a two-body ladder operator, $\frac12 [12:12]$; a three-body linked one, $[12:13]$; and a four-body unlinked one $\frac14 [12:34]$.
The four-body operator is the product of the two independent operators, which give asymmetry factor of $1/2$ individually and $1/4$ in total.
These configurations can be visualized in terms of the diagrammatic representation shown in Fig.~\ref{fig:FF}.
Each configuration has a maximum number $n$ in the square brackets, which leads to the $n$-body operator.
This kind of cluster expansion is systematically performed in any power of the multiple products of the correlation functions \cite{myo15}.

The kinetic energy operator $T$ is treated as one-body term except for the c.m. term $T_{\rm cm}$, which has already been given in Eq.~(\ref{eq:CM}).
The configuration of the one-body operator $T=\sum_i^A t_i$ is expressed as [1] with a single number in the square brackets.
We take the case of $F^\dagger T F $. This correlated operator is expanded into nine diagrams 
from $[12:1:12]$ to $\frac{1}{4}[12:3:45]$, where the single numbers in the middle represent the one-body kinetic operator. 
The corresponding diagrams are displayed in Fig.~\ref{fig:FGF}.

For the two-body interaction $V$, we take the case of $F^\dagger V F $, which is expanded into 16 diagrams 
from $\frac12 [12:12:12]$ to $\frac{1}{8}[12:34:56]$. The corresponding diagrams are displayed in Fig.~\ref{fig:FFF}.
The same scheme of the cluster expansion limited to the single correlation function is presented using shell-model basis states \cite{bishop92,bishop98}.

In the double TOAMD given in Eq.~(\ref{eq:TOAMD2}), we consider the double products of the correlation functions in the wave function.
Among the various kinds of correlated operators, we show the cases of the correlated operators of norm $F^\dagger F^\dagger FF$, one-body kinetic energy $F^\dagger F^\dagger T FF$ and two-body interaction $F^\dagger F^\dagger V FF$
as follows:
\begin{itemize}
\setlength{\itemsep}{6pt}
\renewcommand{\labelitemi}{$\bullet$}
\item $F^\dagger F^\dagger FF$~~~is expanded from $\displaystyle \frac12 [12:12:12:12]$       to $\displaystyle \frac{1}{16}[12:34:56:78]$.
\item $F^\dagger F^\dagger T FF$ is expanded from $\displaystyle [12:12:1:12:12]$             to $\displaystyle \frac{1}{16}[12:34:5:67:89]$.
\item $F^\dagger F^\dagger V FF$ is expanded from $\displaystyle \frac{1}{2}[12:12:12:12:12]$ to $\displaystyle \frac{1}{32}[12:34:56:78:9\,10]$. 
\end{itemize}
The numbers of diagrams of many-body operators arising from the correlated norm, $T$ and $V$ are listed in Table \ref{tab:FFFFF}.
It is necessary to take all the resulting operators in order to retain the variational principle for TOAMD.
In general, multiple products of many correlation functions produce a large number of many-body operators in the cluster expansion.
Among these operators, the higher-body terms require larger calculation costs to obtain their matrix elements numerically, which often occurs for larger mass nuclei.

\begin{table}[t]
\caption{Numbers of diagrams of many-body operators in the cluster expansion of $F^\dagger F^\dagger FF$ (norm), $F^\dagger F^\dagger TFF$ and $F^\dagger F^\dagger VFF$ appearing in double TOAMD.}
\label{tab:FFFFF}
\centering
\begin{tabular}{c|ccccccccc}
\hline
$n$-body &~~~2~~~~&~~~3~~~~&~~~4~~~~&~~~5~~~~&~~~6~~~~&~~~7~~~~&~~~8~~~~&~~~9~~~~&~~~10~~~~\\ \hline \hline
norm     &  1  &  13  &   46   &  47   &  25   &   6   &   1  &  --  & --  \\ \hline
$T$      &  1  &  40  &  183   & 259   & 163   &  55   &  10  &   1  & --  \\ \hline
$V$      &  1  &  40  &  295   & 587   & 516   & 235   &  65  &  10  &  1  \\ \hline
\end{tabular}
\end{table}

\subsection{Energy variation}
The present TOAMD wave function has two kinds of variational functions, the AMD wave function $\Phi_{\rm AMD}$ and the correlation functions $F_D$ and $F_S$.
We determine these functions using the Ritz variational principle with respect to the total energy in TOAMD as $\delta E=0$ in Eq.~(\ref{eq:E_TOAMD}).
For $\Phi_{\rm AMD}$, the centroid positions of the Gaussian wave packet of each nucleon $\{\vec D_i\}$ for $i=1,\ldots,A$ in Eq.~(\ref{eq:Gauss}) are variationally obtained using the cooling method \cite{kanada03}.

The radial forms of $F_D$ and $F_S$ are optimized in four spin--isospin channels to minimize the total energy $E$.
We adopt the Gaussian expansion method to express the pair functions $f^{t}_{D}(r)$ in Eq.~(\ref{eq:Fd}) and $f^{t,s}_{S}(r)$ in Eq.~(\ref{eq:Fs}) as follows:
\begin{eqnarray}
   f^t_D(r)
&=& \sum_{n=1}^{N_G} C^t_n \, e^{-a^t_n r^2}~,\qquad
   f^{t,s}_S(r)
~=~ \sum_{n=1}^{N_G} C^{t,s}_n\, e^{-a^{t,s}_n r^2}~,
   \label{eq:cr}
\end{eqnarray}
where $a^t_n$, $a^{t,s}_n$, $C^t_n$ and $C^{t,s}_n$ are variational parameters.
We take the number of Gaussian functions $N_G=7$, in which we get converging solutions.
For the Gaussian ranges $a^t_n$, $a^{t,s}_n$, we search for their optimized values in a wide range to cover the spatial correlation.
The coefficients $C^t_n$ and $C^{t,s}_n$ are linear parameters in the single correlation terms of TOAMD.
They are determined variationally by diagonalizing the Hamiltonian matrix.
For the double products of the correlation function such as the $F_D F_S$ term in Eq.~(\ref{eq:TOAMD2}), the products of the two Gaussian functions in Eq. (\ref{eq:cr}) become the basis functions and, 
correspondingly, the products of $C^t_n$ and $C^{t,s}_n$ are the variational parameters.

We express the TOAMD wave function in the form of a linear combination using the coefficients of the Gaussian expansion in the correlation functions:
\begin{eqnarray}
   \Phi_{\rm TOAMD}
&=& \sum_{\alpha=0} \tilde{C}_\alpha\,  \Phi_{{\rm TOAMD},\alpha} \,,
   \label{eq:linear}
   \\
   H_{\alpha,\beta}
&=& \langle\Phi_{{\rm TOAMD},\alpha} |H|\Phi_{{\rm TOAMD},\beta}\rangle
~=~ \langle\Phi_{\rm AMD} |\tilde{H}_{\alpha,\beta}|\Phi_{\rm AMD}\rangle \,,
   \nonumber\\
   N_{\alpha,\beta}
&=& \langle\Phi_{{\rm TOAMD},\alpha} |\Phi_{{\rm TOAMD},\beta}\rangle
~=~ \langle\Phi_{\rm AMD} |\tilde{N}_{\alpha,\beta}|\Phi_{\rm AMD}\rangle \,,
   \label{eq:HN}
\end{eqnarray}
where the labels $\alpha$ and $\beta$ are the set of the Gaussian index $n$ and the quantum numbers of spin $s$ and isospin $t$ for two nucleons in the correlation function.
The Hamiltonian and norm matrix elements are $H_{\alpha,\beta}$ and $N_{\alpha,\beta}$, respectively.
For the AMD wave function $\Phi_{\rm AMD}$, we give it the labels $\alpha=0$ and $\beta=0$, and the coefficient $\tilde C_0$ in the linear combination.
The corresponding operators are given as $\tilde{H}_{0,0}=H$ and $\tilde{N}_{0,0}=1$.
For the basis states with a single correlation function, $\tilde C_\alpha$ indicates $C_n^t$ and $C_n^{t,s}$, given in Eq.~(\ref{eq:cr}).

For the double products of the correlation functions, these components are treated as single basis states with the two kinds of correlation functions.
The corresponding expansion coefficients are denoted by $\tilde{C}_\alpha$, which can be $C_n^t C_{n'}^{t'}$, $C_n^t C_{n'}^{t',s}$, and $C_n^{t,s} C_{n'}^{t',s'}$.
Here $\alpha$ includes information on two kinds of correlation functions.
It is noted that the coefficient $\tilde{C}_\alpha$ determined in the calculation cannot be inversely decomposed into $C_n^t$ and $C_n^{t,s}$.
Finally, we solve the following eigenvalue problem to determine the total energy $E$ and all the coefficients $\tilde{C}_\alpha$ in Eq.~(\ref{eq:linear}).
\begin{eqnarray}
   \sum_{\beta=0} \left( H_{\alpha,\beta} - E\, N_{\alpha,\beta} \right) \tilde{C}_\beta &=&0.
   \label{eq:eigen}
\end{eqnarray}

We briefly explain the calculation procedure of the Hamiltonian matrix elements in TOAMD, which are reduced to the matrix elements of the correlated Hamiltonian using the AMD wave function
in Eqs.~(\ref{eq:E_TOAMD}) and (\ref{eq:HN}).
We express the $NN$ interaction $V$ as a sum of Gaussians, similarly to the correlation function $F$.
The correlated operators $\tilde{H}$ and $\tilde{N}$ involve the products of $F$ and $V$. 
After the cluster expansion of $\tilde{H}$ and $\tilde{N}$ to the many-body operators,
the resulting many-body operators include various combinations of the interparticle coordinates in the exponent of the Gaussian. 
This structure of the particle coordinates in the many-body operators makes it difficult to analytically evaluate the matrix elements in general.
In the present approach, we use the Fourier transformation of the Gaussians in $F$ and $V$ \cite{myo15,goto79}.
This transformation decomposes the square of the interparticle coordinates $\vec r_{ij}^{\,2}$ in the exponent into the product of the plane waves, each having single particle coordinate $\vec r_i$ and $\vec r_j$.
In the momentum space, the matrix elements of the many-body operators result in the products of the single-particle matrix elements of the plane waves.
Using the single-particle matrix elements in AMD, we perform multiple integration of the associated momenta in the last step and obtain the matrix elements of TOAMD.
We explain the above procedure in Appendix \ref{TOAMD_ME}. Several typical cases are given in Ref. \cite{myo15}.

\section{Results}\label{sec:results}

\subsection{Central interaction} 
In TOAMD, the central-type correlation function $F_S$ is introduced to express the central-type correlation including the short-range repulsion.
We investigate the applicability of TOAMD to the central interaction case only using the $F_S$ and $F_S F_S$ terms in the double TOAMD wave function in Eq (\ref{eq:TOAMD2}).
We choose the Malfliet--Tjon V (MT-V) of $NN$ interaction \cite{malfliet69,zabolitzky82}, which gives a strong short-range repulsion and a Yukawa-type tail.
The MT-V potential is explicitly defined as
\begin{eqnarray}
   v(r)
&=& 1458.05\ \frac{e^{-3.11 r}}{r} - 578.09\ \frac{e^{-1.55r}}{r} \,,
   \label{eq:MTV}
\end{eqnarray}
in units of MeV and $r$ in units of fm.
The essential results of TOAMD for MT-V are given in Ref. \cite{myo17} for $s$-shell nuclei in comparison with UCOM.
Hence, we show the additional results in this paper.

\begin{figure}[t]
\includegraphics[width=7.3cm,clip]{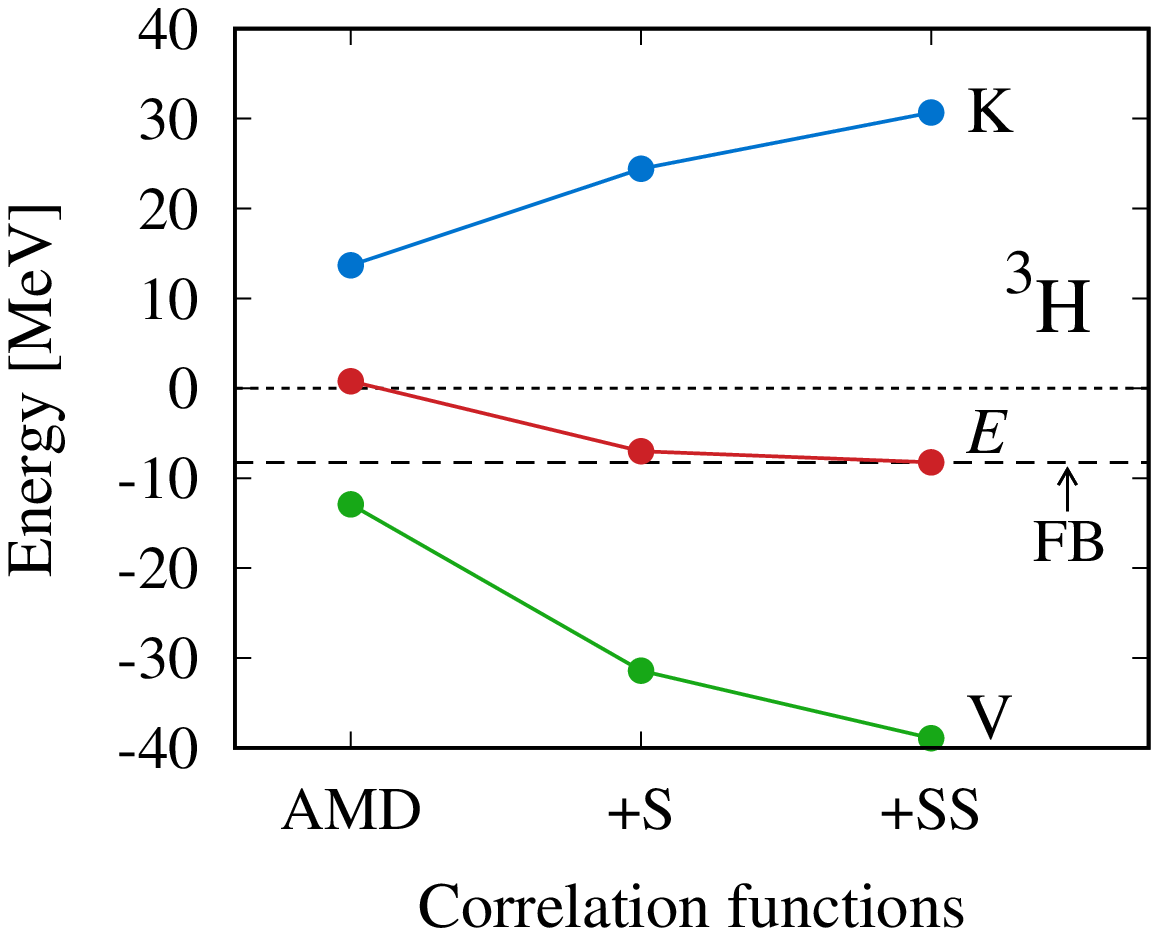}~~~~
\includegraphics[width=7.3cm,clip]{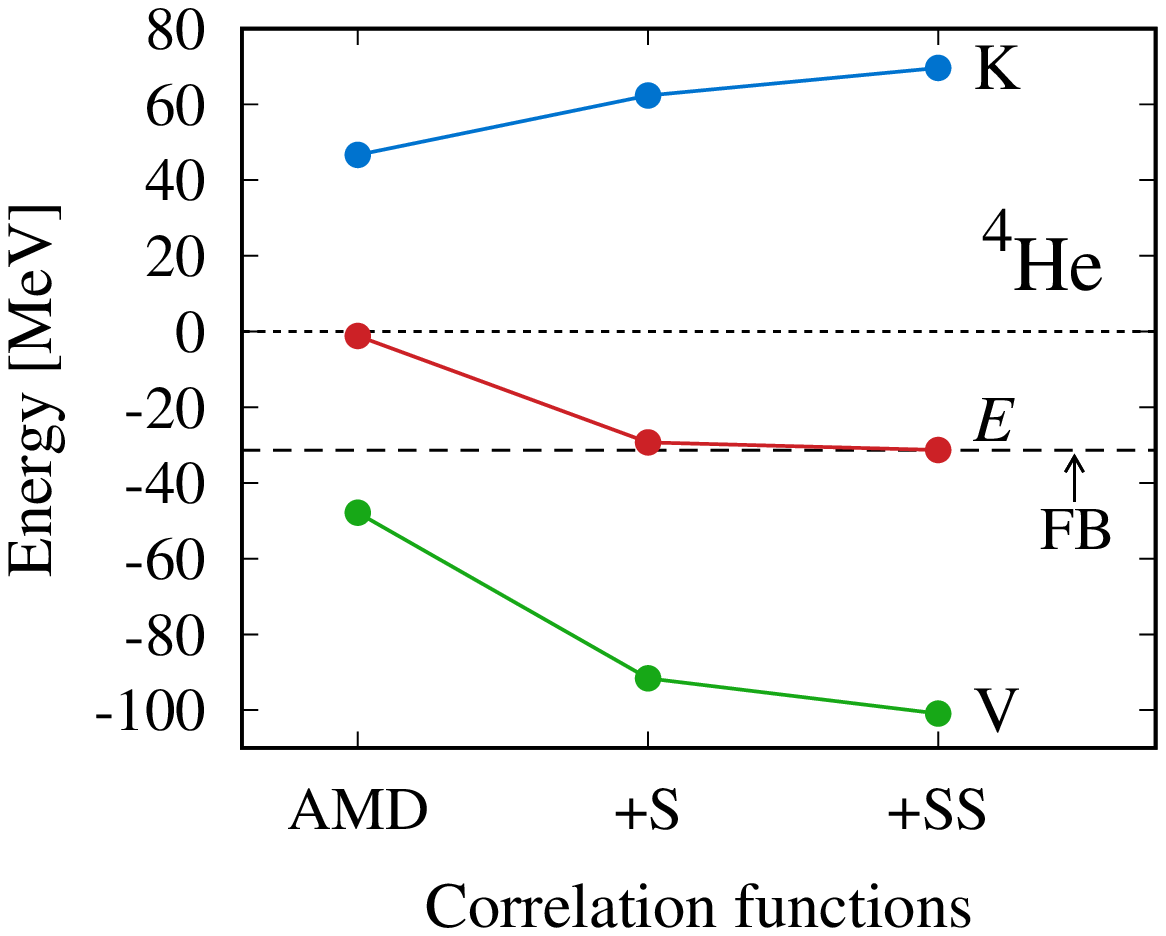}
\caption{
Total energy $E$ and the Hamiltonian components of $^3$H (left) and $^4$He (right) with the MT-V potential by adding each term of TOAMD successively. 
The symbols K and V indicate the kinetic and interaction energies, respectively.
Dashed lines represent the total energies of few-body calculations (FB).
}
\label{fig:ene_MTV}
\end{figure}

In the AMD wave function, we optimize the range parameter $\nu$, which is $0.11$ fm$^{-2}$ for $^3$H and $0.25$ fm$^{-2}$ for $^4$He in the results of the energy minimization.
We also obtain that $\vec D_i=0$ for all nucleons in two nuclei in the single TOAMD calculation. This indicates that the $s$-wave configuration is favored as the AMD wave function.
We keep this condition throughout the analysis with MT-V.
In Fig. \ref{fig:ene_MTV}, we show the results of TOAMD for two nuclei.
We start from the results of AMD and further add the correlation terms successively. 
Here we simply denote the correlation function $F_S$ as ``S''. The symbol +S indicates the wave function of $(1+F_S)\times \Phi_{\rm AMD}$ for single TOAMD,
and the symbol +SS indicates the wave function of $(1+F_S+F_S F_S)\times \Phi_{\rm AMD}$ for double TOAMD.
We can see a nice convergence of energies in double TOAMD to the few-body calculations for two nuclei.
In Fig. \ref{fig:ene_MTV}, at each correlation-addition step, the magnitudes of the kinetic and interaction energies increase together and their cancellation becomes large.
The increase of the kinetic energies indicates that the short-range correlations arising from $F_S$ increase at each step.

In Table \ref{tab:MTV}, we summarize the results of double TOAMD compared with other theories.
It is found that the energies of $^3$H and $^4$He obtained in TOAMD nicely reproduce the few-body results.
In addition, the energies in TOAMD become lower than those of variational Monte Carlo (VMC) \cite{carlson81} for two nuclei.
This result indicates that the accuracy of TOAMD is beyond that of VMC from the variational point of view. 
In VMC, they employ a Jastrow-type correlation function in which the two-body correlation function is multiplied for every pair of nuclei.
This approach is widely used for treating many-body theory in various fields.
In this calculation, they essentially assume the common radial form of the correlation function for any pair. On the other hand, in TOAMD, we take the power series expansion of the correlation functions,
and each correlation function at each order is treated independently and determined variationally with respect to the total wave function.
This indicates that we can optimize the correlation functions of every term fully in TOAMD.
We shall discuss later the advantage of the independent treatment of the correlation functions using the bare $NN$ interaction AV8\,$^\prime$.

\begin{table}[t]
\begin{center}
\caption{Energies of $^3$H($\frac12^+$) and $^4$He ($0^+$) with the MT-V potential in units of MeV in comparison with other theories.}
\label{tab:MTV} 
\begin{tabular}{c|ccccc}
\noalign{\hrule height 0.5pt}
        & VMC~\cite{carlson81} & GFMC~\cite{zabolitzky82} & Few-body~\cite{varga95} &~TOAMD (double) \\
\noalign{\hrule height 0.5pt}
$^3$H   &~~$-8.22(2)$          &~$-8.26(1)$               &~~$-8.25$~~           &~$-8.24$        \\
$^4$He  &~$-31.19(5)$          &~$-31.3(2)$               &~$-31.36$~~           & $-31.28$       \\
\noalign{\hrule height 0.5pt}
\end{tabular}
\end{center}
\end{table}

\subsection{AV8\,$^\prime$ potential}

We discuss the results of TOAMD with the AV8\,$^\prime$ potential for $^3$H and $^4$He. 
This part is the extended analysis reported in Ref. \cite{myo16}.
We report two kinds of the results based on single and double TOAMD.
In Table \ref{tab:nu}, we list the range parameter $\nu$ of $\Phi_{\rm AMD}$ in the TOAMD wave function determined at the single and double levels.
Similar to the case of MT-V central potential, we obtain $\vec D_i=0$ for all nucleons in two nuclei, indicating the $s$-wave configuration of AMD for two nuclei \cite{myo16}. 

In Table \ref{tab:AV8}, we show the results of TOAMD, successively adding the correlation term step-by-step, where we use the $\nu$-value optimized in double TOAMD.
Here we use the labels D and S, indicating the correlation functions $F_D$ and $F_S$, respectively. 
The symbol +D is the result of single TOAMD with a wave function of $(1+F_S+F_D)\times \Phi_{\rm AMD}$ in Eq.~(\ref{eq:TOAMD1}). 
The symbol +SS is the result of adding the $F_S F_S$ term as $(1+F_S+F_D+F_S F_S) \times \Phi_{\rm AMD}$.
The symbol +DD is the full calculation in double TOAMD defined in Eq.~(\ref{eq:TOAMD2}).
The components of $F_S F_D$ and $F_D F_S$ haven an almost identical effect on the solutions; hence, the their results are combined and denoted by +SD+DS.

We confirm that, starting from AMD, which gives positive energies, the energies of $^3$H and $^4$He are converged successively to the results of GFMC.
This behavior is also shown in Figs.\,\ref{fig:ene_3H_fix} and \ref{fig:ene_4He_fix}, explained later.
The energy difference between GFMC and double TOAMD is 80 keV for $^3$H and 1.2 MeV for $^4$He.

\begin{table}[t]
\begin{center}
\caption{Range parameter $\nu$ optimized in TOAMD for $^3$H ($\frac12^+$) and $^4$He ($0^+$) with AV8$^\prime$ potential in units of fm$^{-2}$.}
\label{tab:nu} 
\begin{tabular}{c|cccc}
\noalign{\hrule height 0.5pt}
TOAMD   &~~Single~~      &~~Double~~ \\
\noalign{\hrule height 0.5pt}
$^3$H   &  $0.14$        & $0.095$  \\
$^4$He  &  $0.17$        & $0.22$   \\
\noalign{\hrule height 0.5pt}
\end{tabular}
\end{center}
\end{table}

\begin{table}[t]
\begin{center}
\caption{Energies of $^3$H ($\frac12^+$) and $^4$He ($0^+$) with the AV8$^\prime$ potential in units of MeV, 
where the range parameters $\nu$ are determined in double TOAMD, listed in Table \ref{tab:nu}.}
\label{tab:AV8} 
\begin{tabular}{c|c|cc|ccc|c}
\noalign{\hrule height 0.5pt}
        &~AMD~    &~~+S~~  & ~~+D~    &~ +SS  ~  &~+SD+DS  &~~+DD ~   &  GFMC \cite{kamada01} \\
\noalign{\hrule height 0.5pt}
$^3$H   & $11.37$ & $2.58$ &~$-4.98$  &~$-6.12$  &~$-7.27$  &~$-7.68$  &~$-7.76$  \\
$^4$He  & $56.42$ & $7.90$ & $-14.74$ & $-18.13$ & $-22.44$ & $-24.74$ & $-25.93$ \\
\noalign{\hrule height 0.5pt}
\end{tabular}
\end{center}
\end{table}

\subsubsection{Single TOAMD}
We first discuss the results of TOAMD with the single correlation function given in Eq.~(\ref{eq:TOAMD1}).
This analysis provides the fundamental properties of the TOAMD wave function.
After that, we proceed to the analysis of double TOAMD.

\begin{figure}[t]
\includegraphics[width=7.7cm,clip]{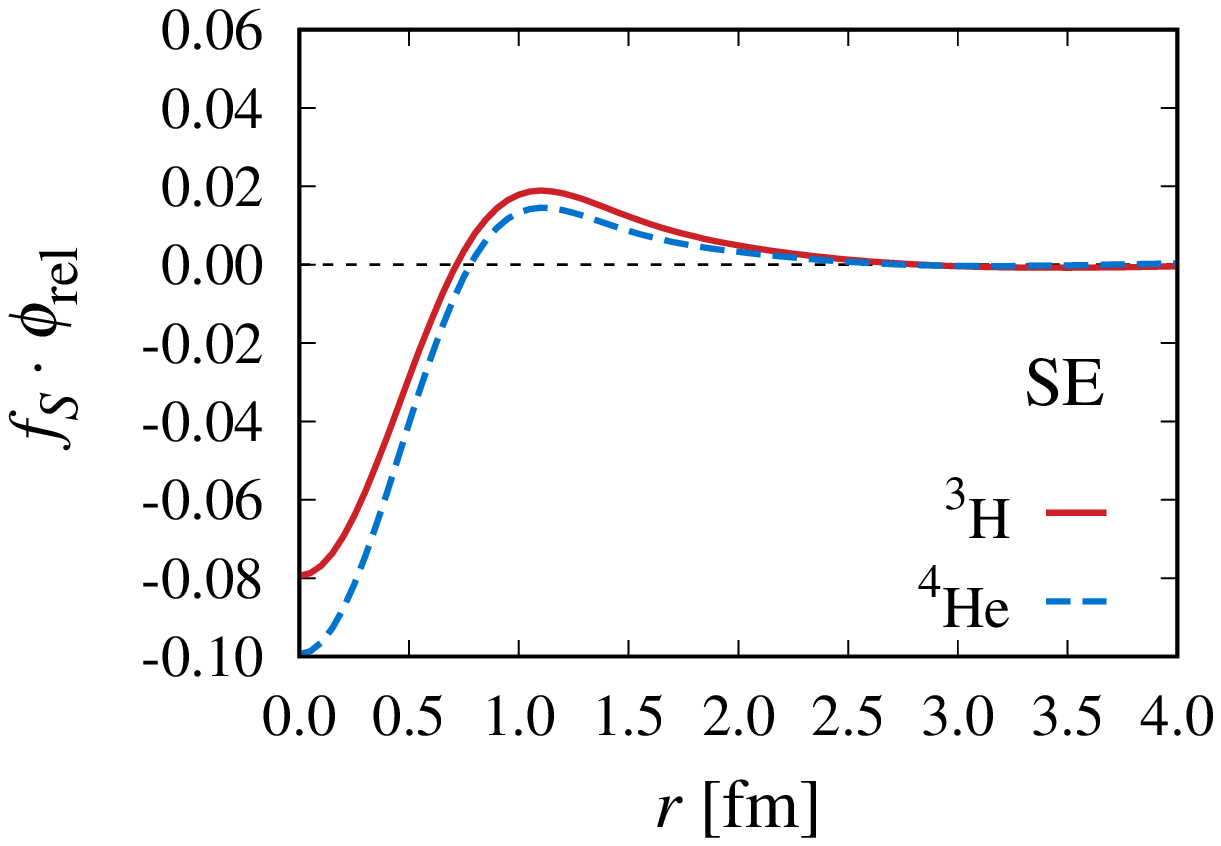}~
\includegraphics[width=7.7cm,clip]{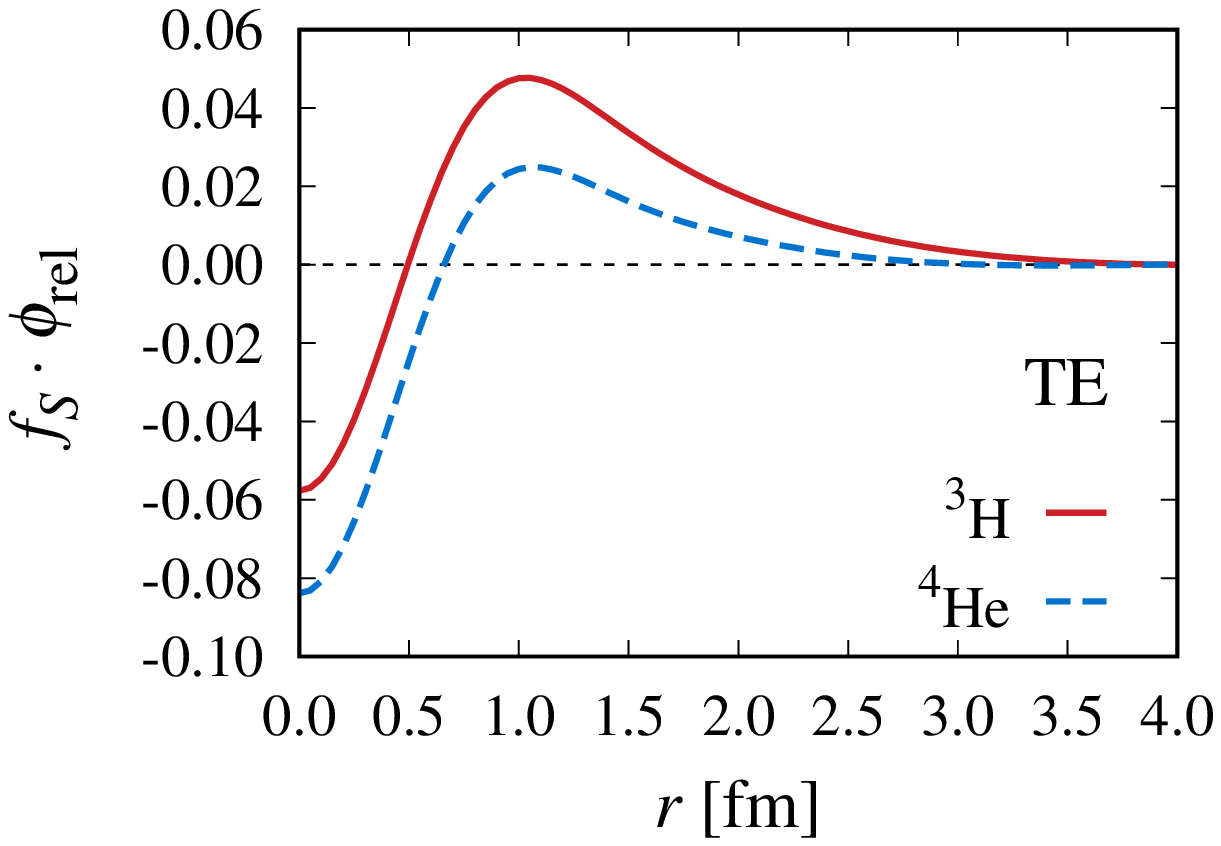}
\caption{Short-range correlation functions $f_S(r)$ of the singlet-even (SE, left) and triplet-even (TE, right) channels in single TOAMD for $^3$H and $^4$He 
multiplied by the relative wave function $\phi_{\rm rel}(r)$ of AMD.}
\label{fig:cr_S}
\end{figure}

\begin{figure}[t]
\centering
\includegraphics[width=7.7cm,clip]{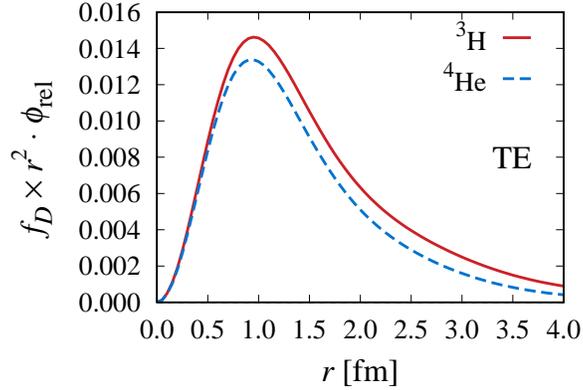}
\caption{Tensor correlation functions $f_D(r)$ of the triplet-even (TE) component in single TOAMD for $^3$H and $^4$He 
multiplied by $r^2$ and the relative wave function of AMD.}
\label{fig:cr_D}
\end{figure}

The energies of $^3$H and $^4$He are $-5.34$ MeV and $-15.68$ MeV, respectively, in single TOAMD with the optimized ranges $\nu$ listed in Table \ref{tab:nu}.
It is interesting to plot the correlation functions $f_S(r)$ for the short-range part and $f_D(r)$ for the tensor part in Eq.~(\ref{eq:cr}), 
which are determined variationally using Gaussian expansion. 
We multiply the relative wave function of two nucleons $\phi_{\rm rel}(\vec r)$ in AMD with the correlation functions, which is physically meaningful.
The relative wave function $\phi_{\rm rel}(\vec r)$ is defined in the form of a Gaussian wave packet as
\begin{eqnarray}
    \phi_{\rm rel}(\vec r)
&=& \left(\frac{2\nu_r}{\pi}\right)^{3/4} e^{-\nu_r(\vec r - \vec D_r)^2}  
    \\
    \nu_r
&=& \frac{\nu}{2}\, ,\qquad
    \vec D_r
~=~ \vec D_1 -\vec D_2\,.
\end{eqnarray}
Here, $\vec D_1$, $\vec D_2$, and $\vec D_r$ are the centroid of the two nucleon positions and thee difference between them, respectively; they are set to zero in the results of variation in TOAMD. 
This result makes $\phi_{\rm rel}$ spherical. In addition only the even channels are considered because of the $s$-wave configuration of AMD. 
The range parameter $\nu_r$ is for the relative wave function of each nucleus.
In Figs.~\ref{fig:cr_S} and \ref{fig:cr_D}, we show the functions $f_S\cdot \phi_{\rm rel}$ and $f_D\cdot \phi_{\rm rel}$.
For short-range correlation, the singlet-even (SE) and triplet-even (TE) channels are shown in Fig.\,\ref{fig:cr_S}. 
They commonly have negative values at short distances below 0.5 fm. This behavior represents a reduction of the short-range amplitude of the nucleon pair from the original AMD wave function.
Beyond 0.5 fm, the functions become positive and show maximum values at about 1.0 fm, corresponding to the intermediate distance.
This indicates that $f_S(r)$ has two characters; one is the short-range correlation to avoid the repulsion in the interaction and the other is the intermediate correlation 
to obtain the attraction from the interaction.
The results of $^3$H and $^4$He exhibit a similar trend.
The reduction of short-range amplitude in SE is larger than that in the case of TE. 
This is related to a feature of the AV8$^\prime$ potential: short-range repulsion of SE is stronger than that of TE \cite{wiringa95}.
The size of the intermediate amplitude in TE is larger than that in SE.
This property is related to the presence of coupling between the $S$- and $D$-wave states via the tensor force in TE to obtain the total energy.

For the tensor correlation in Fig.\,\ref{fig:cr_D}, the TE channel is shown, where the term of $r^2$ is multiplied in relation to the definition of $F_D$ in Eq.~(\ref{eq:Fd}). 
The distribution shows a peak at around 1.0 fm for the two nuclei, which agrees with the peak position in the TE channel of $f_S(r)$ as shown in Fig.~\ref{fig:cr_S}.
This agreement of the peak positions makes the tensor coupling increase in the TE channel.
From the behavior of the correlation functions, the ranges of $f_S(r)$ and $f_D(r)$ are not short.
This property increases the contributions of the many-body terms of the correlated operators in the cluster expansion beyond the two-body approximation. 
This point will be discussed later with the Hamiltonian components.

\begin{figure}[t]
\centering
\includegraphics[width=8.0cm,clip]{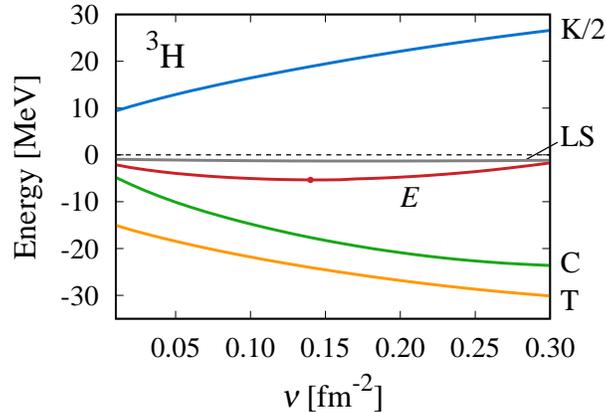}
\caption{Energy surface of $^3$H as a function of the range parameter $\nu$ in single TOAMD with AV8$^\prime$.
Total energy $E$ and the Hamiltonian components are shown.
For the kinetic energy (K), a half-value is shown with the symbol K/2. The symbols C, T, and LS indicate the central, tensor, and $LS$ forces, respectively.
}
\label{fig:ene_3H_F}
\end{figure}

In Fig. \ref{fig:ene_3H_F}, we plot the $\nu$-dependence of the total energy and the Hamiltonian components for $^3$H.
Here, small $\nu$ represents the spatially extended AMD wave function and large $\nu$ represents the spatially compact case.
It is noted that the $\nu$-dependence corresponds to the $\hbar\omega$-dependence in the shell-model prescription,
because of the $s$-wave configuration of the AMD wave function.
We clearly confirm the saturation behavior of the total energy $E$. Among the Hamiltonian components, the contribution of the tensor force is larger than that of the central force in the AV8$^\prime$ potential
for any range of $\nu$.
As the range of $\nu$ becomes large, the contributions of the interaction energy and also the kinetic energy increase, which causes a large cancellation between them to obtain the total energy.

\begin{figure}[t]
\includegraphics[width=7.5cm,clip]{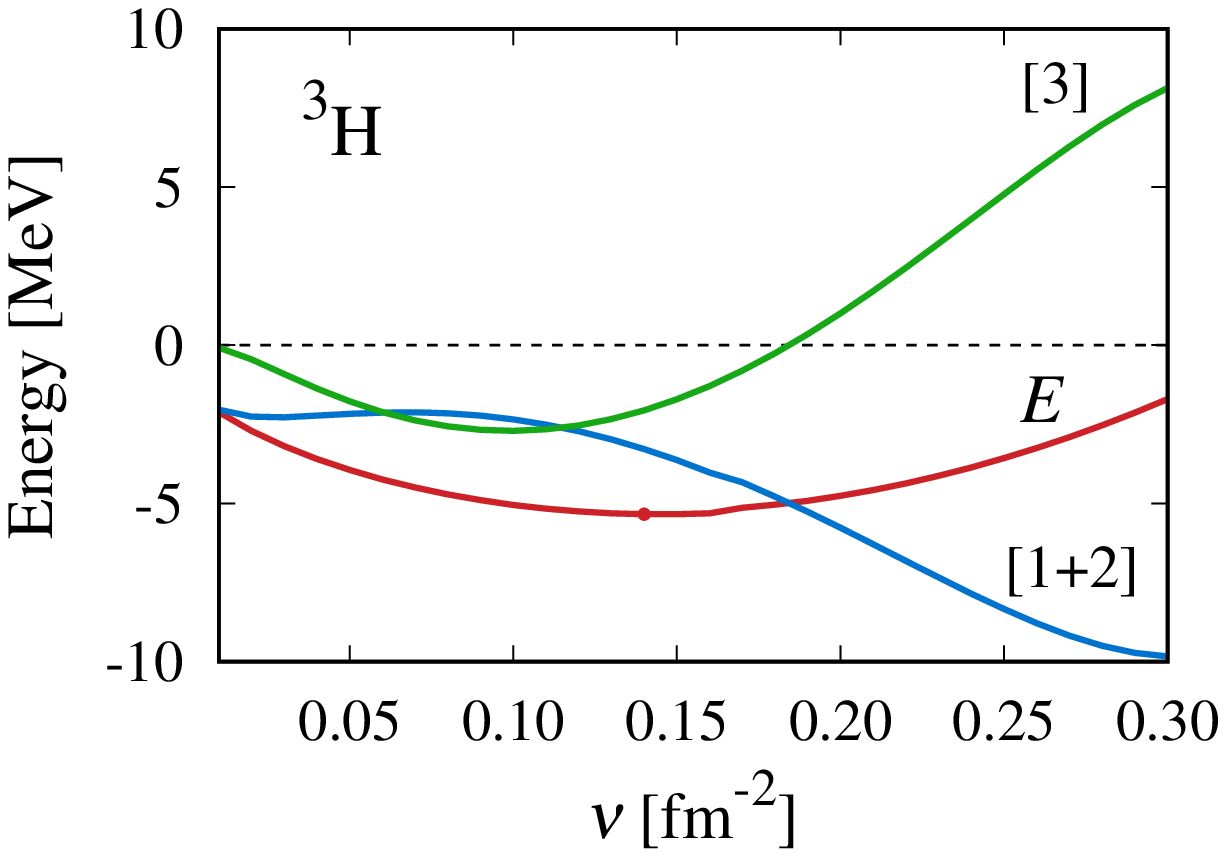}~~
\includegraphics[width=7.5cm,clip]{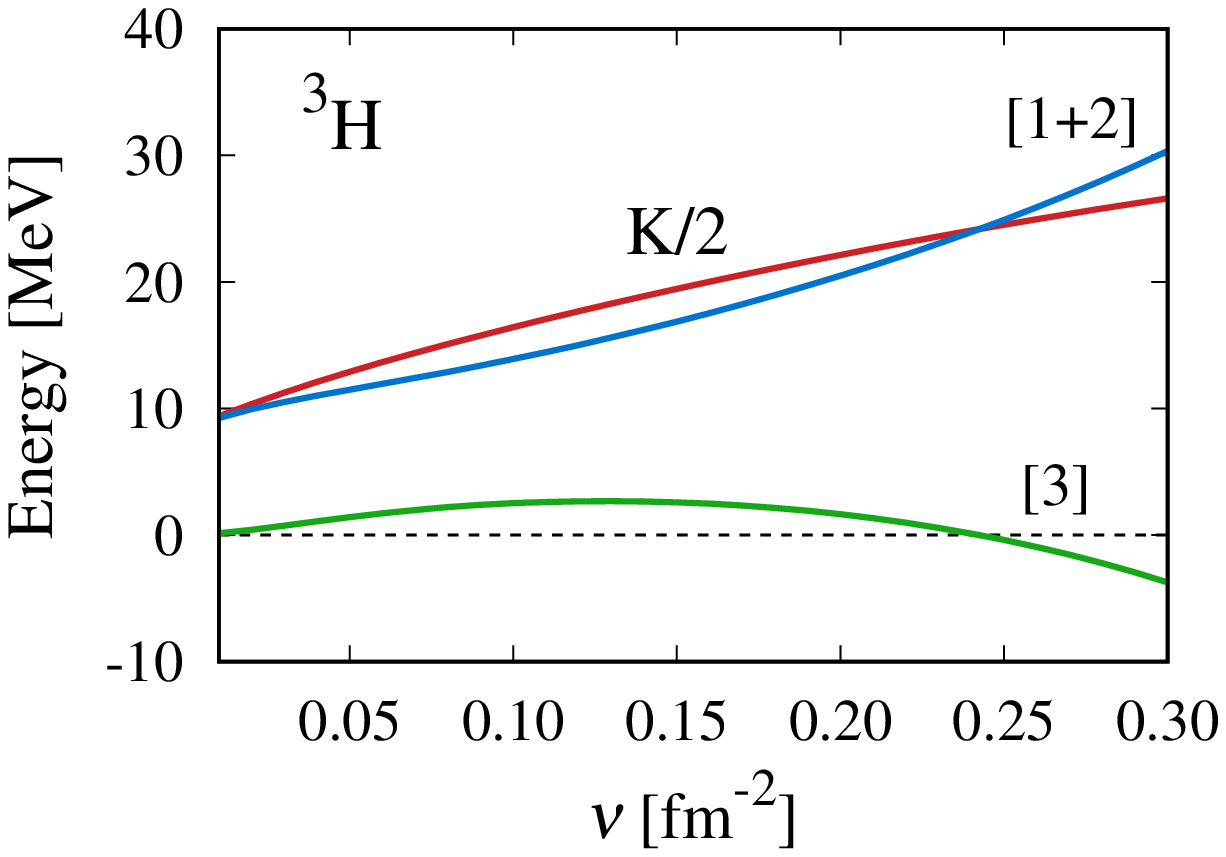}\\
\includegraphics[width=7.5cm,clip]{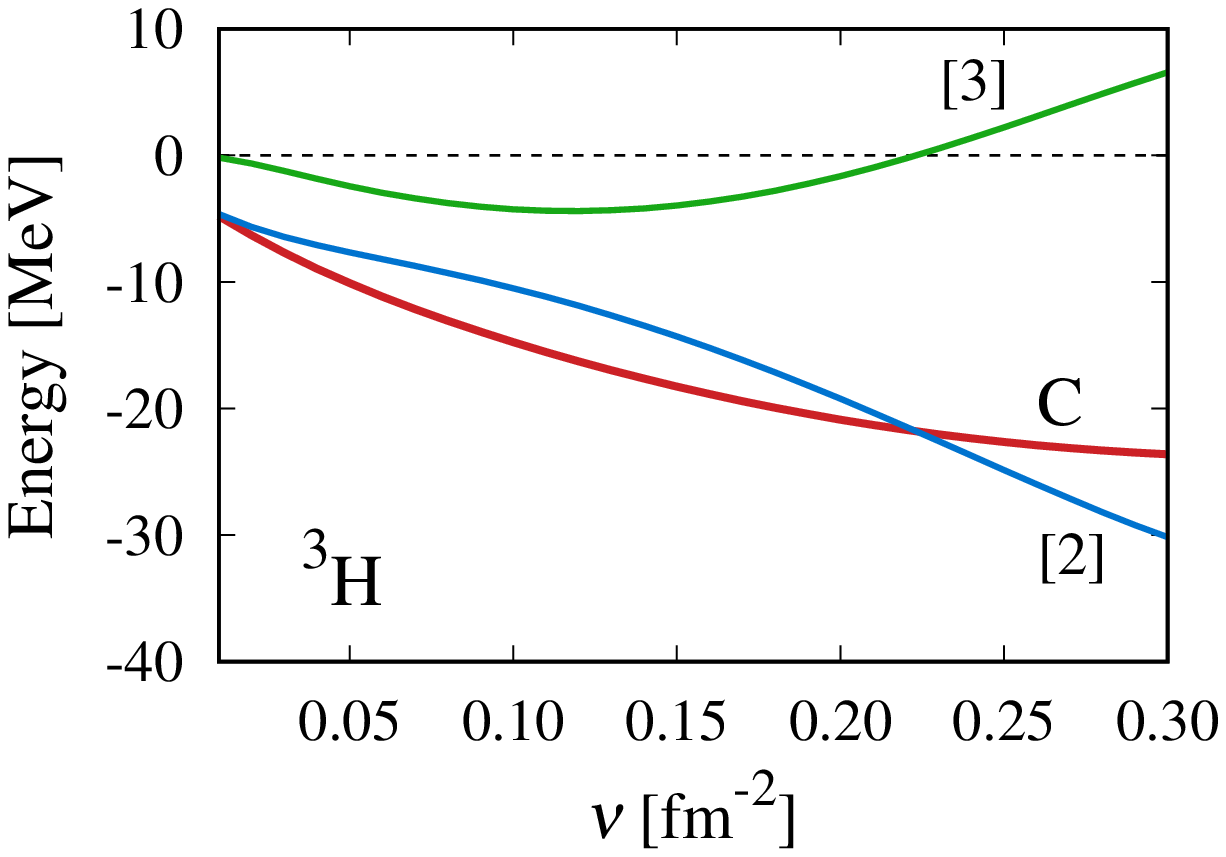}~~
\includegraphics[width=7.5cm,clip]{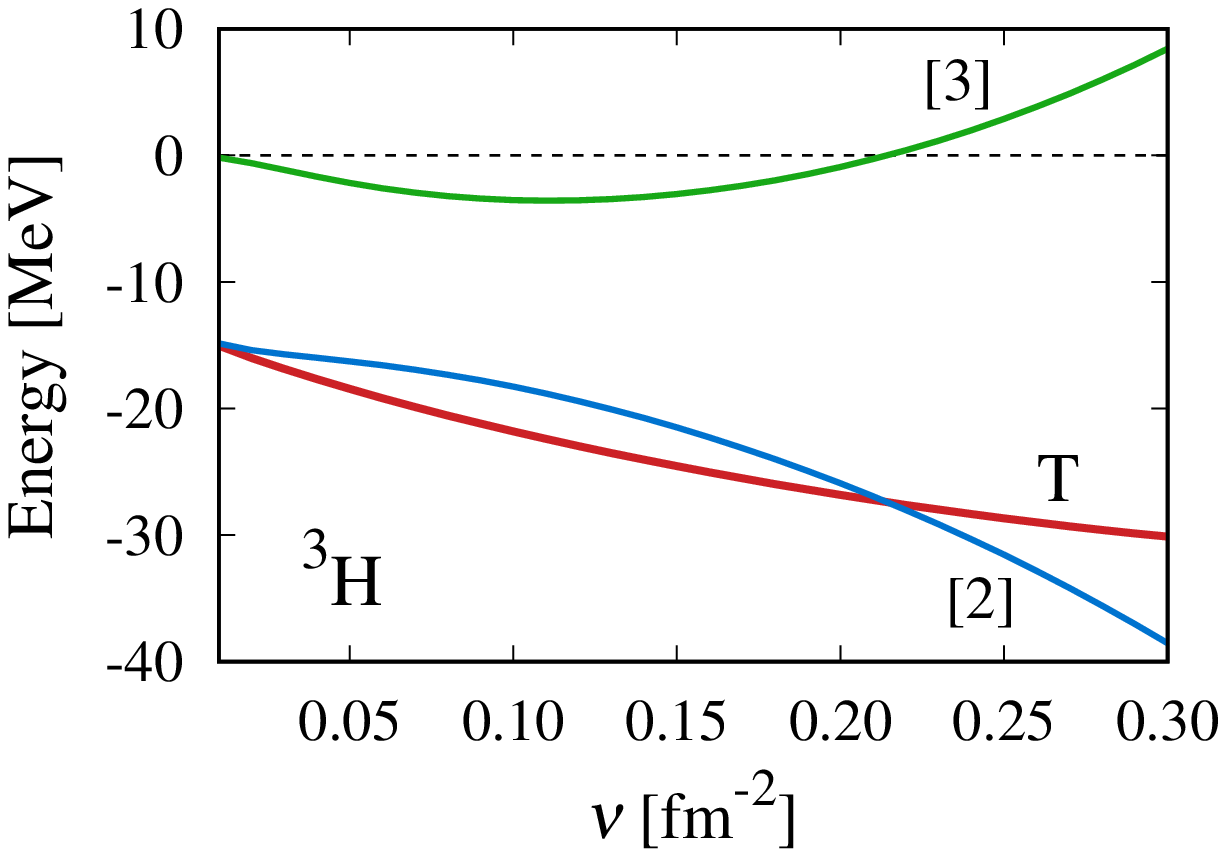}
\caption{
Many-body terms of the total energy ($E$, upper-left panel), the kinetic (K, upper right), central (C, lower left), and tensor (T, lower right) components of $^3$H in single TOAMD.
The symbols [2] and [3] indicate the two-body and three-body terms in the cluster expansion, respectively.
For the kinetic energy, its half-value (K/2) is shown. The term of [1+2] in the total and kinetic energies represents a sum of the one-body kinetic energy and the two-body term.}
\label{fig:ham_3H_F}
\end{figure}

We discuss the explicit role of the many-body operators in the cluster expansion of the correlated Hamiltonian.
In Fig.\,\ref{fig:ham_3H_F}, we decompose the total energy and the Hamiltonian components into many-body terms,
while we sum up all the cluster components in the correlated norm part appearing in the denominator in Eq.~(\ref{eq:E_TOAMD}).
The symbols [2] and [3] represent the two-body and three-body cluster terms of the correlated operators, respectively.
The symbol [1+2] appearing in the total and kinetic energies represents the sum of the one-body kinetic energy $T$, in which the c.m. term is subtracted and the two-body term.
This [1+2] term corresponds to the calculation under the two-body approximation of the transformed Hamiltonian such as UCOM.
For the total energy $E$, it is found that the [1+2] term cannot provide the saturation behavior around the energy minimum point at $\nu=0.14$ fm$^{-2}$.
On the other hand, the three-body term [3] shows the saturation at small $\nu$ value of 0.10 fm$^{-2}$ and becomes repulsive at large $\nu$.
The sum of all the cluster terms gives the saturation behavior of the total energy as a function of $\nu$.
This decomposition indicates the important role of the three-body term in the correlated Hamiltonian.

We show the contribution of each Hamiltonian component of $^3$H in Fig. \ref{fig:ham_3H_F}.
For the kinetic energy (K), central (C) and tensor (T) forces, we decompose their correlated operators into each cluster term.
For the kinetic energy, the terms [1+2] and [3] are shown.
It is found that the [1+2] term makes a dominant contribution and the thee-body term [3] makes a the small contribution in scale. 
For central and tensor forces, the dominant contributions come from the two-body term [2], which is common,
and the three-body term [3] gives the small contribution in scale. These results are similar to those of the kinetic energy.
From the decomposition, all those up to the two-body term make the main contribution in the correlated Hamiltonian, but these are largely canceled out between the kinetic and interaction energies.
The three-body terms of each Hamiltonian component commonly give the small value in magnitude.
In total, owing to the large cancellation between the kinetic and interaction energies,
the three-body term [3] can be compared to the [1+2] term in the total energy as shown in Fig. \ref{fig:ham_3H_F}.

\begin{figure}[t]
\centering
\includegraphics[width=8.0cm,clip]{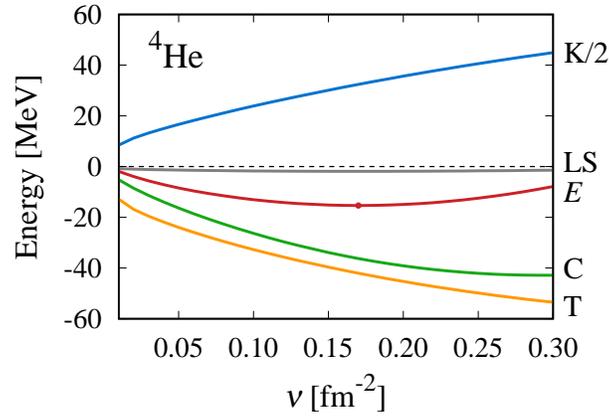}
\caption{Energy surface of $^4$He as function of range parameter $\nu$ in single TOAMD with AV8$^\prime$.
Notations are the same as used in Fig. \ref{fig:ene_3H_F}.
}
\label{fig:ene_4He_F}
\end{figure}
\begin{figure}[t]
\includegraphics[width=7.5cm,clip]{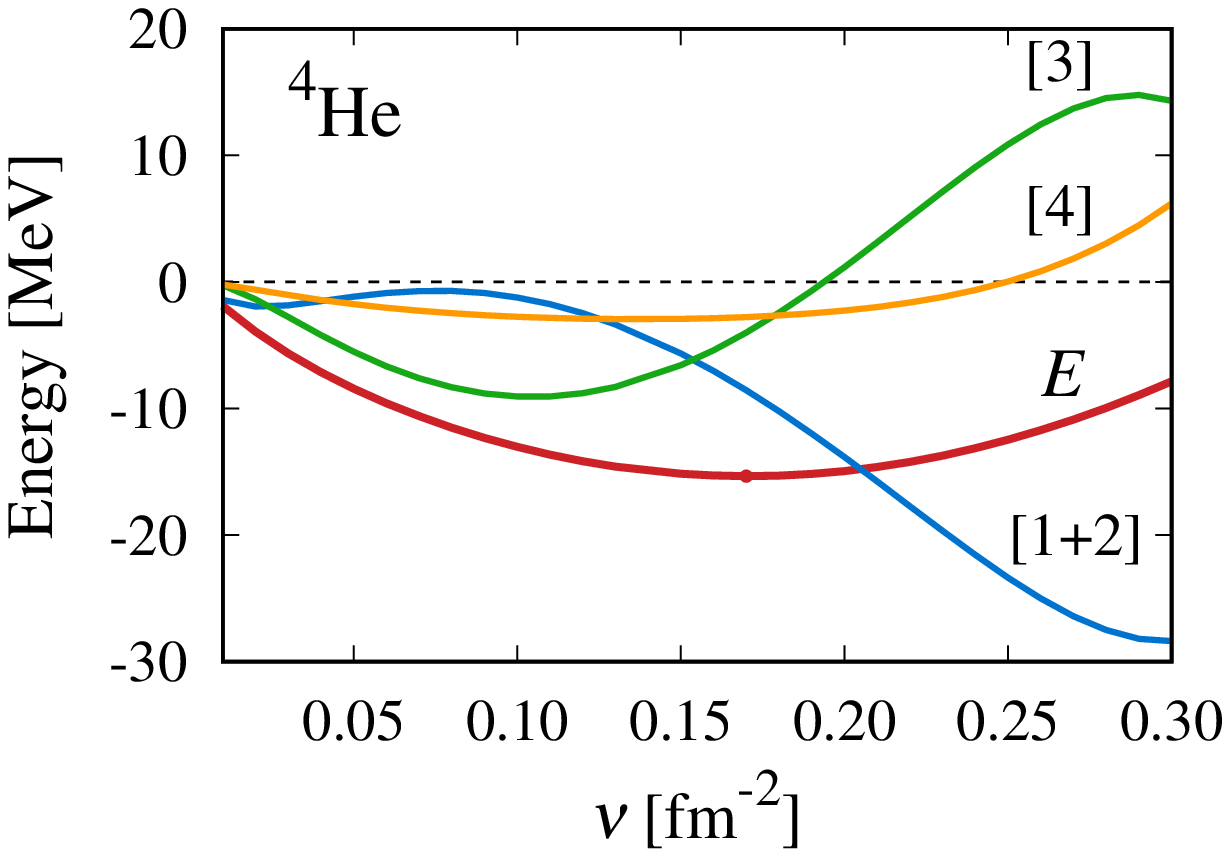}~~
\includegraphics[width=7.5cm,clip]{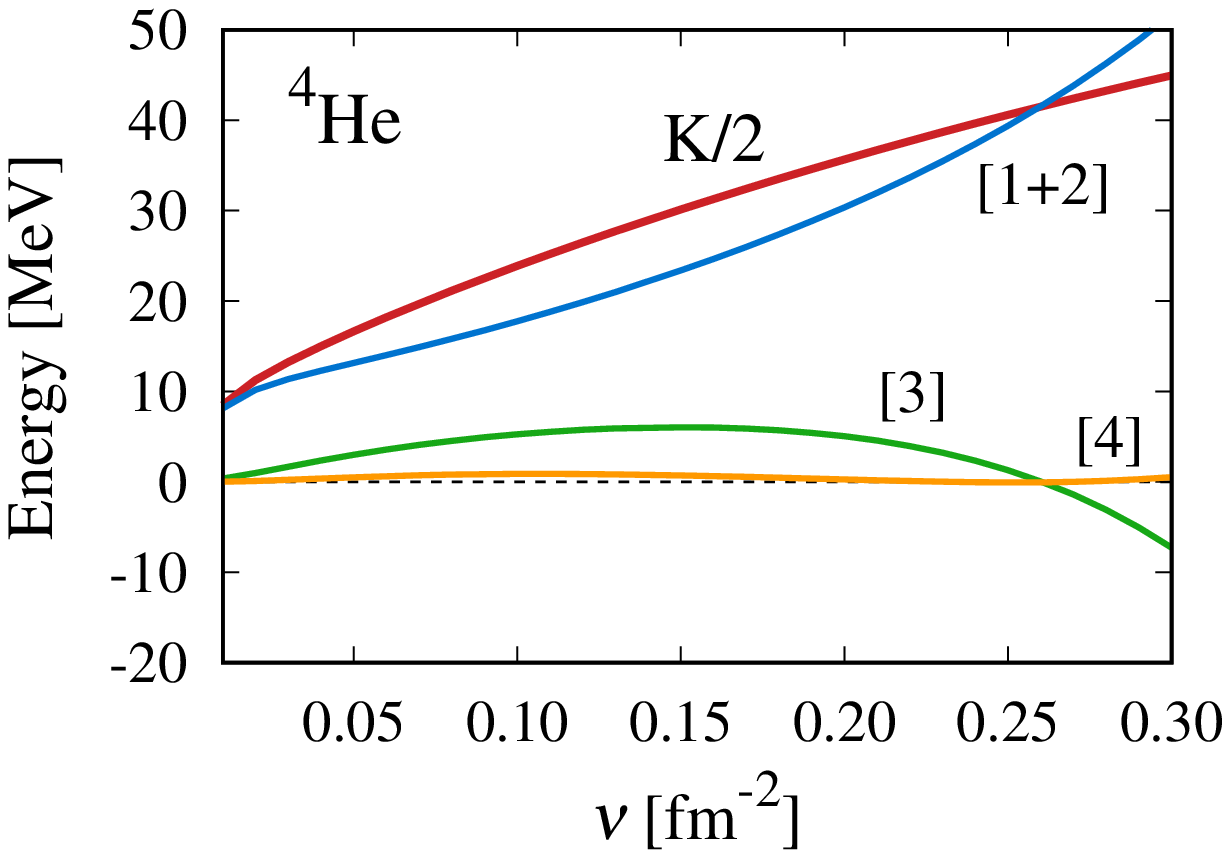}\\
\includegraphics[width=7.5cm,clip]{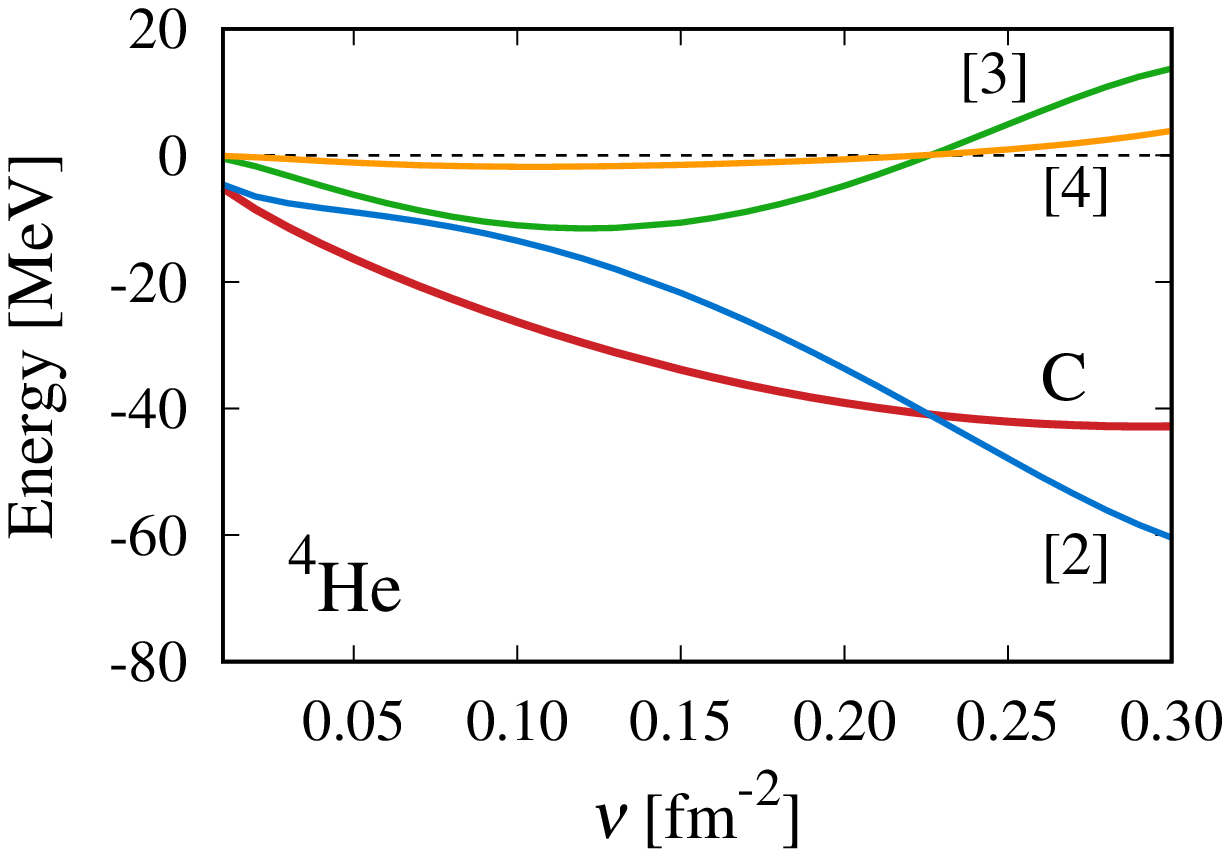}~~
\includegraphics[width=7.5cm,clip]{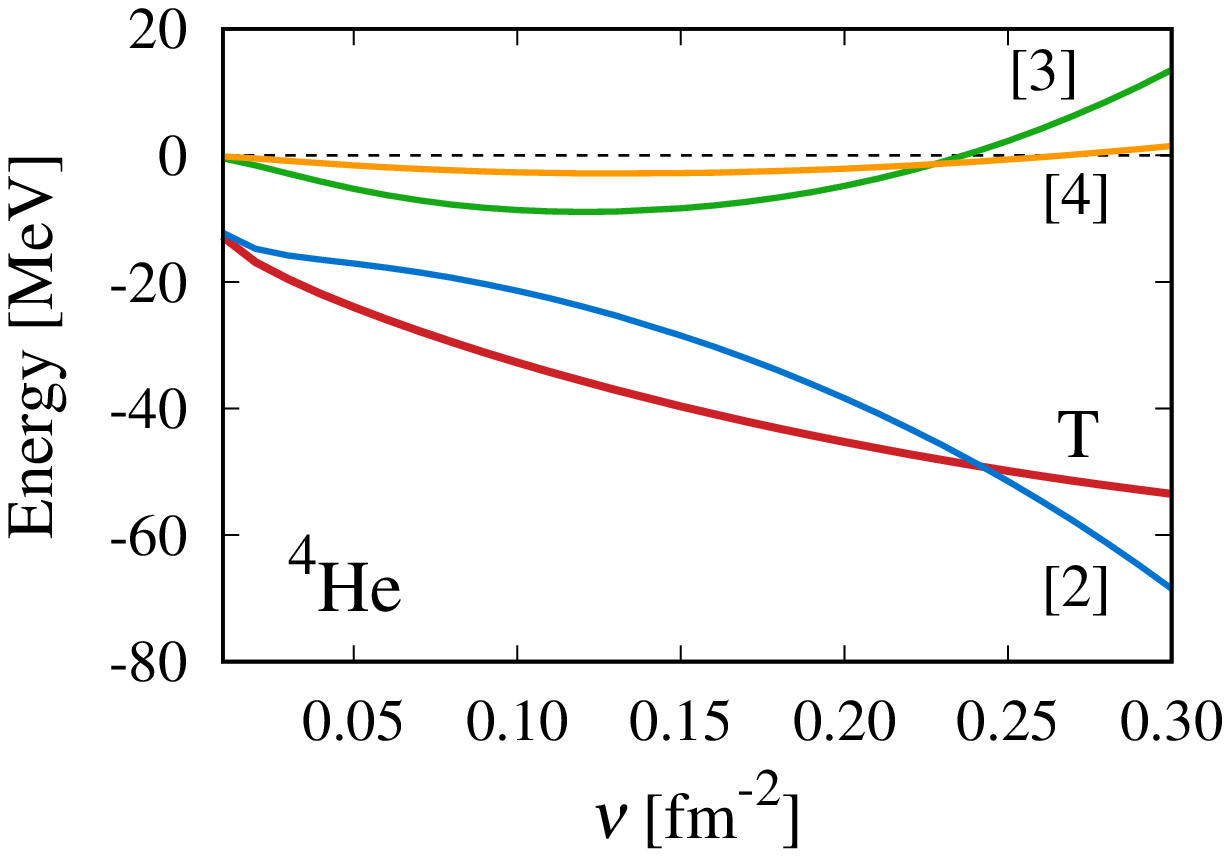}
\caption{Many-body terms of the total energy ($E$) and the Hamiltonian components (K, C, and T) of $^4$He in single TOAMD.
Notations are the same as used in Fig. \ref{fig:ham_3H_F}.
}
\label{fig:ham_4He_F}
\end{figure}

We proceed to the analysis of $^4$He.
In Fig.\,\ref{fig:ene_4He_F}, we plot the $\nu$-dependence of the Hamiltonian components of $^4$He.
We clearly confirm the saturation behavior for the total energy, where the optimized range $\nu=0.17$ fm$^{-2}$ is larger than the value of $^3$H of 0.14 fm$^{-2}$ in Fig. \ref{fig:ene_3H_F}, 
because $^4$He is spatially more compact than $^3$H.
Among the Hamiltonian components, the contribution of the tensor force is larger than that of the central force in the AV8$^\prime$ potential, similar to the results of $^3$H.

In Fig.\,\ref{fig:ham_4He_F}, we decompose the total energy $E$ into the many-body terms of the correlated Hamiltonian in the cluster expansion as [1+2], [3], and the newly four-body term [4].
It is found that the [1+2] term cannot present the saturation behavior at around the optimized range $\nu$ in the total energy.
The three-body [3] and four-body [4] terms provide the contributions, which are not so small.
Owing to the presence of higher-body terms beyond the two-body one, the total energy has a proper energy minimum, as any variational method should provide.

We see the cluster contributions of each Hamiltonian component of $^4$He in Fig. \ref{fig:ham_4He_F} up to the four-body cluster term. 
For the kinetic energy, the [1+2] term makes the main contribution, and the three-body term [3] makes a smaller contribution in scale than that of the [1+2] term.
The four-body term [4] makes a much smaller contribution in scale than the values of [1+2] and [3].
For the central and tensor forces, the dominant contributions commonly come from the two-body term [2], 
the three-body term [3] makes a small contribution in scale, and the four-body term [4] makes a much smaller contribution than the others.
From these results, all those up to the two-body term make the main contribution in the correlated Hamiltonian. 
The three- and four-body terms of each Hamiltonian component commonly give small values.
The higher-body terms tend to make smaller contributions in each component of the correlated Hamiltonian.
On the other hand, the contributions from the kinetic and interaction energies are largely canceled out, which results in the small total energy.
In total, the sum of the three-body and four-body terms in the total energy $E$ is comparable to the sum of the two-body term, as shown in Fig.\,\ref{fig:ham_4He_F}.

In the summary of single TOAMD, we confirm the importance of many-body cluster terms in the correlated Hamiltonian.
This result indicates the inevitable role of many-body operators in the correlated Hamiltonian to describe the energy minimum required from the variational point of view.
A similar result is obtained with the Brueckner--Bethe--Goldstone approach for nuclear matter \cite{fukukawa15}, 
in which the three-body correlations induced by the $G$-matrix are necessary to explain the energy saturation in nuclear matter.
In TOAMD, the higher-body terms tend to make a smaller contribution in the cluster expansion of the correlated operators,
but we cannot ignore these terms because of the large cancellation between the kinetic and interaction energies.
The contribution of the higher-body terms is related to the spatial range of the correlation functions $F_S$ and $F_D$, which is not short in TOAMD, as shown in Figs.~\ref{fig:cr_S} and \ref{fig:cr_D}.
The present analysis also indicates that the two-body approximation should be treated carefully in relation to the range of the correlation functions in many-body theory.

\subsubsection{Double TOAMD}
\begin{figure}[t]
\centering
\includegraphics[width=8.5cm,clip]{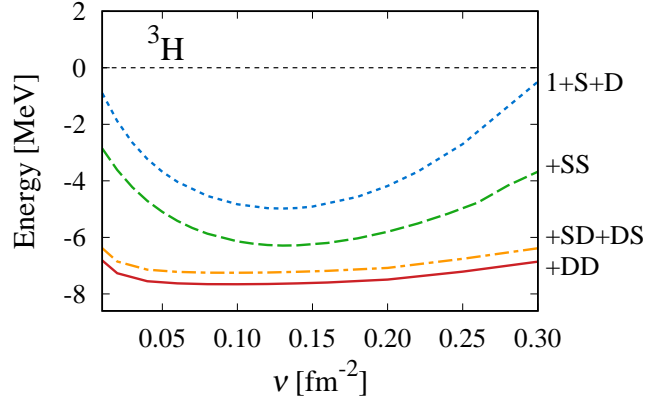}
\caption{Surface of total energy $E$ of $^3$H using AV8$^\prime$ as a function of the range parameter $\nu$ in TOAMD with a single correlation function (1+S+D, dotted line) 
and double correlation functions, adding terms successively from $F_S F_S$ (+SS) to $F_D F_D$ (+DD) (dashed, dash-dotted, and solid lines, respectively).
}
\label{fig:ene_3H_FF_nu}
\end{figure}

\begin{table}[t]
\caption{Energy and the Hamiltonian components of $^3$H and $^4$He in doubled TOAMD. Units are in MeV.
The values in parentheses are obtained in the few-body calculations for $^3$H and GFMC for $^4$He \cite{suzuki08,kamada01}.}
\label{tab:energy_FF}
\centering
\begin{tabular}{cccccc}
        & Energy    & Kinetic   & Central    & Tensor     & $LS$      \\ \hline
$^3$H   &  $-$7.68  &  47.21    & $-$22.44   & $-$30.60   & $-$1.86   \\ 
        & ($-$7.76) & (47.57)   &($-$22.49)  &($-$30.84)  &($-$2.00)  \\ \hline 
$^4$He  & $-$24.74  &  97.06    & $-$53.12   & $-$64.84   & $-$3.83   \\
        &($-$25.93) & (102.3)   &($-$55.05)  &($-$68.05)  &($-$4.75)  \\ \hline
\end{tabular}
\end{table}

We discuss the results of double TOAMD.
We include the double products of the correlation functions in TOAMD given in Eq.\,(\ref{eq:TOAMD2}).
We keep the nucleon positions $\vec D_i=0$ for all nucleons, which are obtained in the results of single TOAMD.

In Fig. \ref{fig:ene_3H_FF_nu}, we show the $\nu$-dependence of the energy of $^3$H in double TOAMD by adding the terms successively.
It is confirmed that the energy surface of $^3$H becomes deeper and flatter by adding the correlation terms in the wave function.
This $\nu$-independence corresponds to the $\hbar \omega$-independence.
The flat property of the energy curve represents the flexibility of the correlation functions, which are optimized as much as possible 
at any range of $\nu$ to include the correlations in the TOAMD wave function. As a result the correlation functions gives a stable solution of TOAMD 
with respect to the range parameter $\nu$.

\begin{figure}[t]
\includegraphics[width=7.7cm,clip]{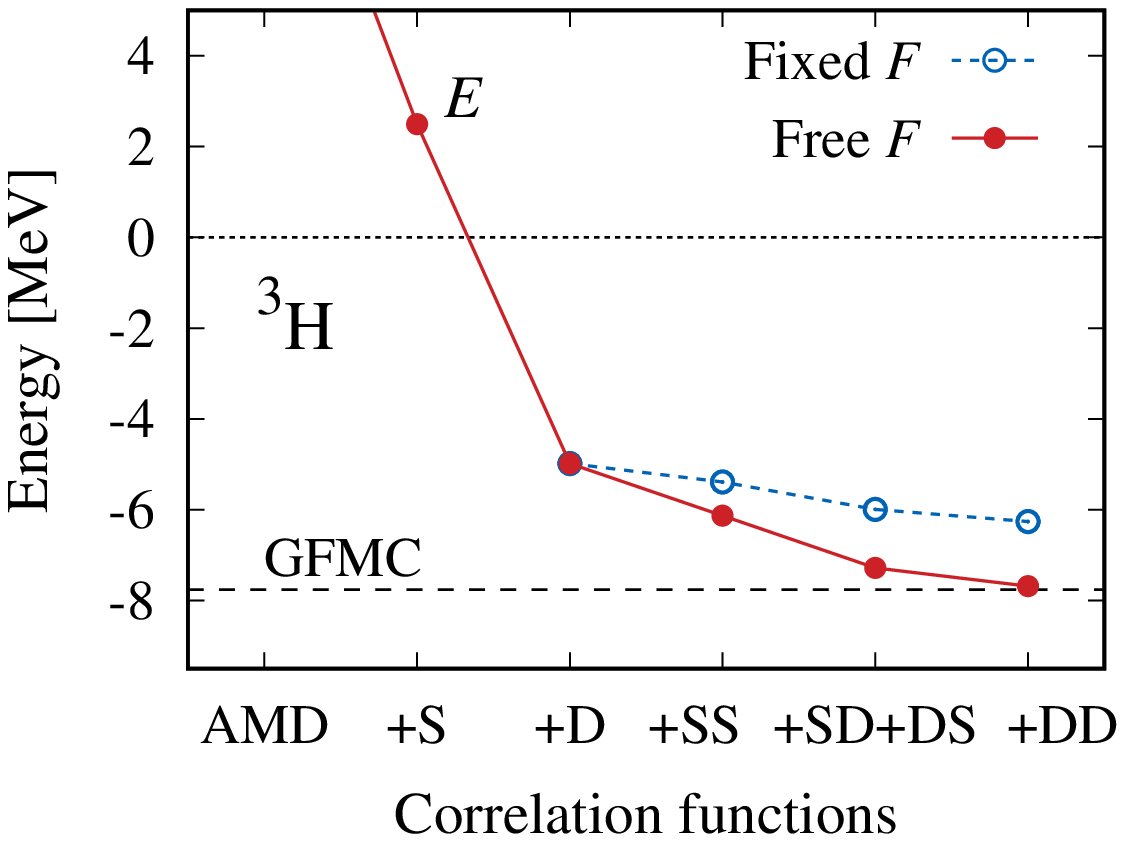}~
\includegraphics[width=7.7cm,clip]{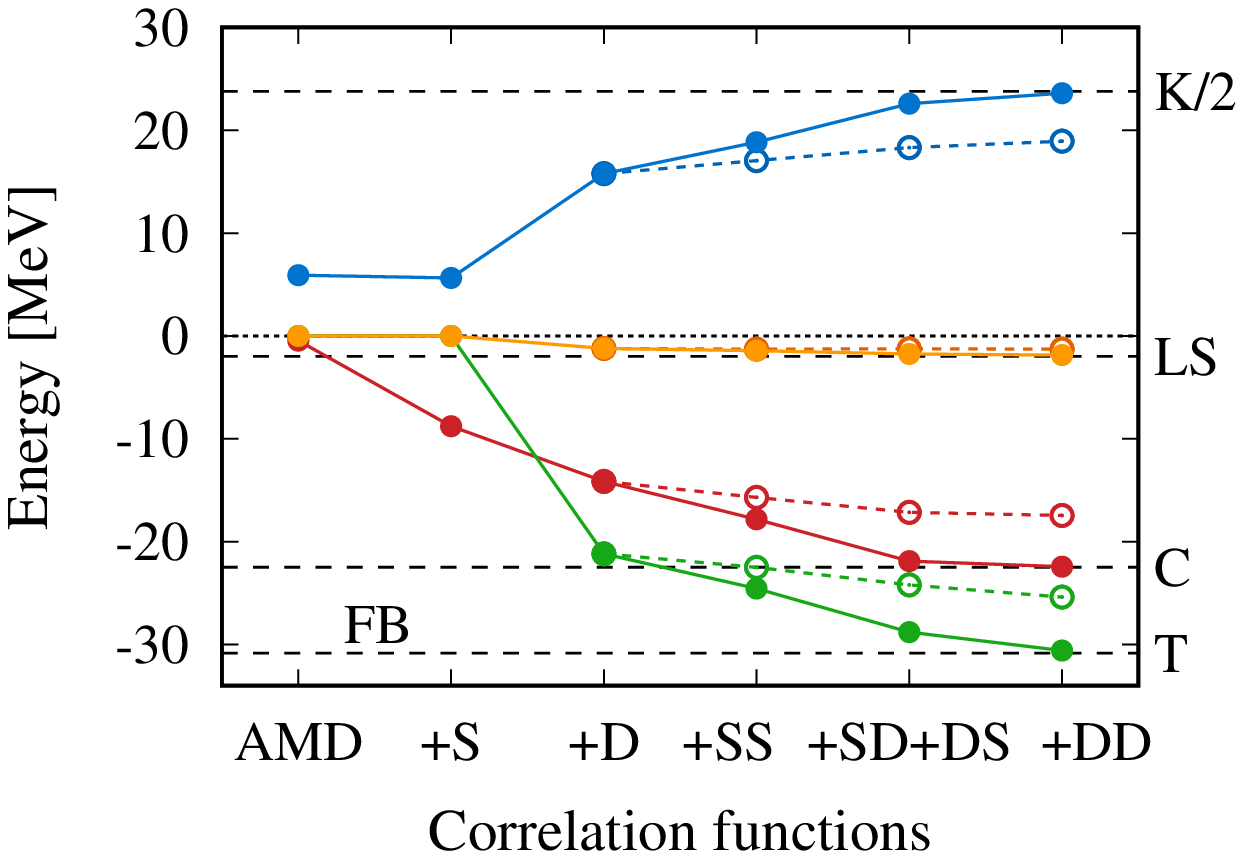}
\caption{(Left) Convergence of total energy $E$ of $^3$H with AV8$^\prime$ by adding each term of TOAMD successively. 
(Right) Hamiltonian components of $^3$H by adding each term of TOAMD successively.
For the kinetic energy, a half-value is shown with the symbol K/2. The symbols C, T, and LS indicate the central, tensor, and $LS$ forces, respectively.
In both figures, solid circles (open circles) indicate the results using the fully optimized (fixed) correlation functions. 
Horizontal dashed lines represent the results of other theories.}
\label{fig:ene_3H_fix}
\end{figure}

In Fig.\,\ref{fig:ene_3H_fix}, we show the total energy $E$ of $^3$H obtained by adding the single and double correlation functions one by one with the solid circles. 
In each calculation, we set the value of $\nu$ to 0.095 fm$^{-2}$, as shown in Table \ref{tab:nu}, but the correlation functions are optimized in each calculation.
The final energy with $F_D F_D$ (+DD) is $-7.68$ MeV, as shown in Table~\ref{tab:AV8}.
The Hamiltonian components are shown in Table \ref{tab:energy_FF}.
We can see the good agreement between TOAMD and other theories for each component.
The matter radius is obtained as 1.746 fm in TOAMD.

Figure \ref{fig:ene_3H_fix} also explains the convergence of the energy, which is the same as that shown in Table~\ref{tab:AV8}.
We confirm the converging behavior of the total-energy curve toward the values of GFMC.
Similarly, we discuss the contributions of the kinetic energy (K), central (C), tensor (T), and $LS$ (LS) forces in Fig.\,\ref{fig:ene_3H_fix}. 
The contributions of the tensor and $LS$ forces appear after adding the tensor correlation (+D), which leads to the mixing of the $D$-wave component.
We can see a nice convergence of each Hamiltonian component toward the results of few-body calculations (FB).
From these results, TOAMD is able to treat the $NN$ interaction explicitly 
owing to the two kinds of correlation functions $F_D$ and $F_S$ and is regarded as a successive variational method for nuclear many-body systems.

\begin{figure}[t]
\includegraphics[width=7.7cm,clip]{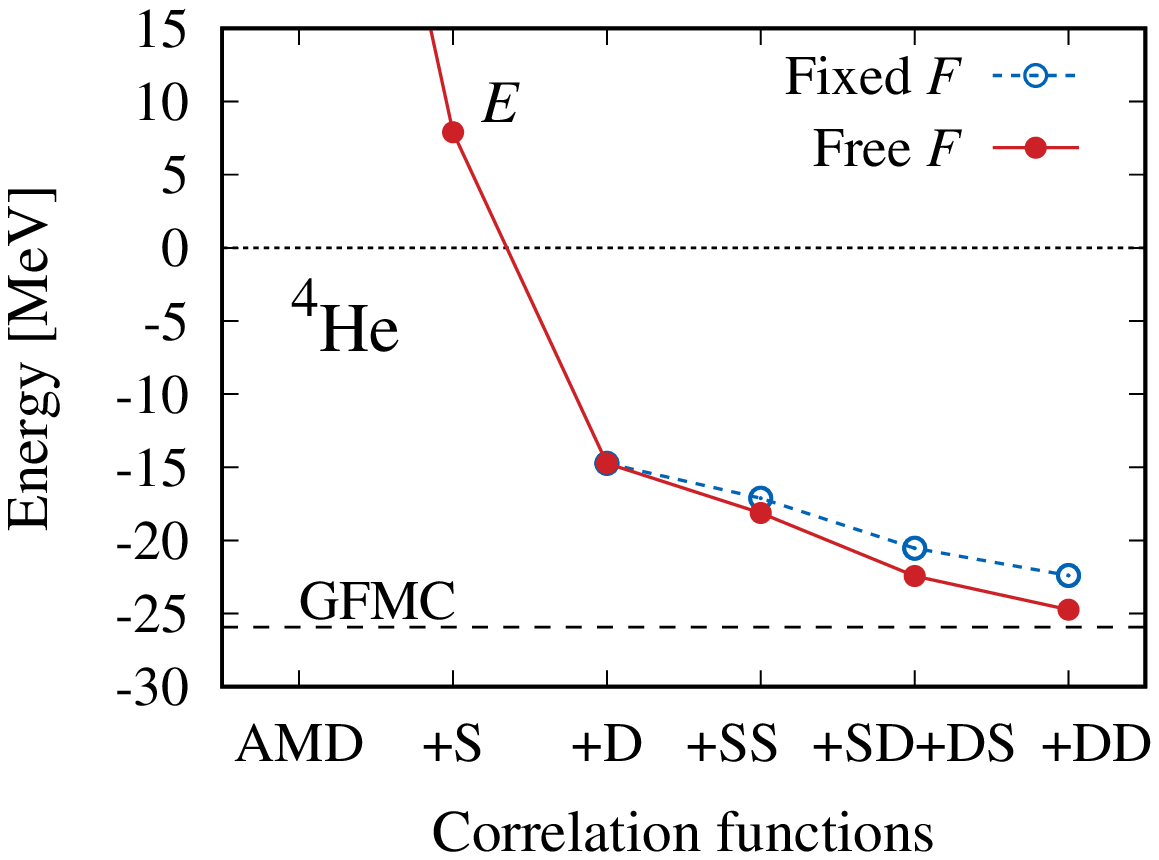}~
\includegraphics[width=7.7cm,clip]{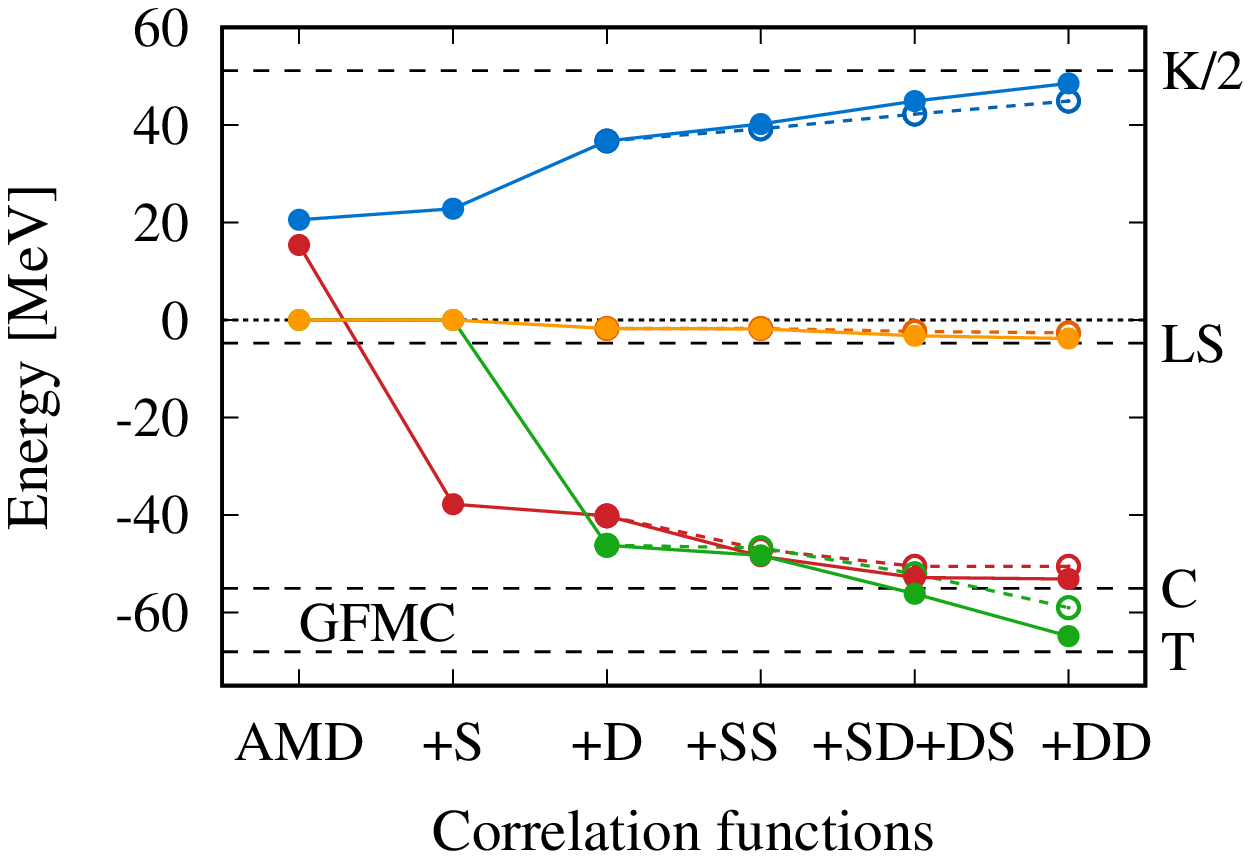}
\caption{
(Left) Convergence of the total energy of $^4$He with AV8$^\prime$ by adding each term of TOAMD successively.  
(Right) Hamiltonian components of $^4$He by adding each term of TOAMD successively.
Notations are the same as used in Fig. \ref{fig:ene_3H_fix}.
}
\label{fig:ene_4He_fix}
\end{figure}

We discuss the case of $^4$He. 
The solid circles in Fig.\,\ref{fig:ene_4He_fix} show the total energy obtained by successively adding the correlation functions in the TOAMD wave function in Eq.~(\ref{eq:TOAMD2}). 
The behavior is very similar to that of $^3$H. We see a good convergence with the correlation functions. 
Finally we obtain $-24.74$ MeV of the energy in double TOAMD. 
The matter radius is obtained as 1.497 fm in TOAMD, reproducing the GFMC value of 1.490 fm \cite{kamada01}. 
Figure\,\ref{fig:ene_4He_fix} also shows the Hamiltonian components.  
Each component of the kinetic energy (K), central (C), tensor (T), and $LS$ (LS) forces shows good convergence with successive addition of the correlation terms, 
but slight deviation from the GFMC results, which suggests the inclusion of the triple products of the correlation functions in TOAMD as a subject of further work.

\begin{figure}[t]
\centering
\includegraphics[width=7.7cm,clip]{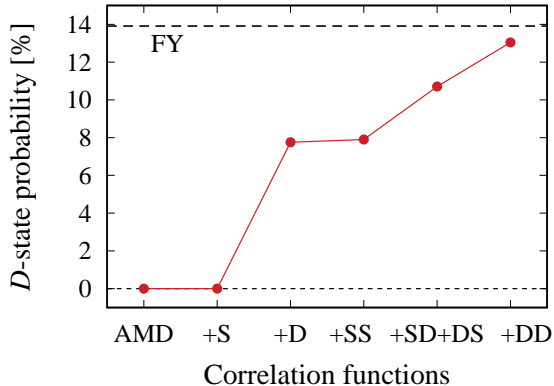}
\caption{$D$-state probability of $^4$He with AV8$^\prime$ by adding each term of TOAMD successively.
The dashed line represents the Faddeev--Yakubovsky (FY) calculation \cite{kamada01}.}
\label{fig:D_state}
\end{figure}

In Fig.\,\ref{fig:D_state}, we show the $D$-state probability of $^4$He.
This quantity is defined as the component of the wave function in which the total orbital angular momentum and total intrinsic spin are both two and they are coupled to be zero.
In double TOAMD, the $D$-state probability is 13.04 which is close to the value of 13.91 obtained in the Faddeev--Yakubovsky (FY) calculation \cite{kamada01}. 
We expect that the small difference will be recovered by adding the higher-order terms in TOAMD such as the triple products of the correlation functions including $F_D$, 
similarly to the Hamiltonian components.

In TOAMD, we emphasize that the correlation functions $F_D$ and $F_S$ are optimized independently in each term of Eq.~(\ref{eq:TOAMD2}).
This is different from the Jastrow approach in which they assume the same functional form of the correlation function for every pair.
It is interesting to examine the effect of the independent treatment of the correlation functions on the solutions of  TOAMD.
For this purpose we perform the following calculation:
First, $F_S$ and $F_D$ are determined in single TOAMD with $(1+F_S+F_D) \times \Phi_{\rm AMD}$.
Second, keeping the functional form of $F_S$ and $F_D$ determined with the Gaussian expansion, we perform the calculation of double TOAMD,
where only the weights of the single and double correlation functions are variational parameters.
The open circles in Figs.\,\ref{fig:ene_3H_fix} and \ref{fig:ene_4He_fix} show the results of these constraint calculations, which we call ``Fixed $F$'' for $^3$He and $^4$He.
At each step of double TOAMD, we can see the energy difference from the original calculation, ``Free $F$''.
Finally this condition provides the energies of $^3$H and $^4$He as $-6.26$ MeV and $-22.40$ MeV, respectively,
giving an energy loss from the full calculation of 1.44 MeV for $^3$H and 2.34 MeV for $^4$He.
These amounts are not small and indicate the importance of the independent optimization of the correlation functions at each term of TOAMD. 
This treatment of the correlation functions also contributes to the rapid energy convergence in TOAMD.
Among the Hamiltonian components in Figs.\,\ref{fig:ene_3H_fix} and \ref{fig:ene_4He_fix},
every component of ``Fixed $F$'' shows a smaller value in magnitude than those of the full calculations.
This result indicates that both $F_D$ and $F_S$ should be optimized in each term of TOAMD.

We also perform the calculation with a different condition from above, in which, after the full calculation of double TOAMD, we extract $F_S$ and $F_D$ from the single correlation function terms.
These $F_S$ and $F_D$ are commonly used in the double correlation terms of TOAMD.
This condition provides $^3$H and $^4$He energies of $-4.95$ MeV and $-21.67$ MeV, respectively, which become worse in comparison with 
the above results in which the correlation functions are fixed in single TOAMD.
These results indicate that, in double TOAMD, the correlation functions included in the single and double correlation terms should be different each other and this property is variationally favored.

Including the double TOAMD components for both the MT-V central potential and the AV8$^\prime$ bare potential, 
the binding energies and the Hamiltonian components of $^3$H and $^4$He are successively converged to values very close to those in the few-body calculations.  
It is found that the accuracy of the results obtained in TOAMD depends on the interactions.
For the difference between the TOAMD and few-body calculations,
the amounts of the difference using AV8$^\prime$, shown in Table \ref{tab:energy_FF}, are larger than those using MT-V, shown in Table \ref{tab:MTV}.
This result suggests that more correlations should be included successively in TOAMD 
in the case of bare interactions, which is possible by increasing the multiple products of the correlation functions including $F_D$.
We shall consider the triple case, such as $F_D F_D F_S$ and $F_D F_S F_S$ in subsequent work; 
this is expected to increase the numerical accuracy of TOAMD using the bare interaction.
The other way to extend TOAMD is the multi-basis representation of the AMD wave function from the single basis case.

\section{Summary}\label{sec:summary}

We have performed a detailed analysis of $s$-shell nuclei with a new variational theory of ''tensor-optimized antisymmetrized molecular dynamics'' (TOAMD) \cite{myo15,myo16,myo17}.
In TOAMD, we introduce two kinds of correlation functions, $F_D$ and $F_S$ for the tensor and short-range correlations, respectively
 to treat the strong interaction directly in the description of nuclei.
We employ AMD as a reference state and multiply the correlation functions by the AMD basis state.
The scheme of TOAMD is extendable by increasing the series of the multiple products of the correlation functions by successive power expansion. 
It is noted that each correlation function included in each term of the multiple products is treated independently
and their functional form can be different. 
This property of TOAMD is different from the ordinary Jastrow method, in which the correlation functions are multiplied by every particle pair in a common form.
The present formulation of TOAMD is applicable to all nuclei with various mass numbers.

In TOAMD, the product of the correlation function and the Hamiltonian is taken into account as the correlated Hamiltonian and is treated in terms of the cluster expansion.
The correlated Hamiltonian becomes a series of many-body operators.
We explicitly manipulate these operators in the calculation of the matrix elements in TOAMD
without any truncation of the higher-body operators beyond the two-body one.
This is an important point to retain the variational principle for the TOAMD wave function.
In this paper, we employ diagrammatic representation to distinguish each diagram obtained in the cluster expansion of the correlated operators.

We have shown the results of $s$-shell nuclei with TOAMD including up to the double products of the correlation functions (double TOAMD).
We employ two kinds of $NN$ interactions of the MT-V central potential with strong short-range repulsion and the AV8$^\prime$ bare potential with tensor and $LS$ forces.
The energies and Hamiltonian components obtained in TOAMD nicely reproduce the few-body results of $^3$H and $^4$He such as GFMC in both interactions.
These results show the power of TOAMD and the efficiency of two correlation functions $F_D$ and $F_S$ to treat the $NN$ interaction.

We show the spatial distributions of the correlation functions $F_D$ and $F_S$, which are found to be not short-range.
In particular $F_S$ has two roles in the reduction of the short-range amplitude of the nucleon pair 
and the attraction at the intermediate range, where the latter is related to $F_D$ via the $S$--$D$ coupling in the triplet-even channel.
We also discuss the role of many-body terms of the correlated Hamiltonian in the cluster expansion, in particular, more than the two-body term.
It is found that the contributions of three-body and four-body terms are smaller than that of the two-body term, but not negligible in each Hamiltonian component of the kinetic and interaction energies.
Owing to the large cancellation between the kinetic and interaction energies, the contribution of the higher-body terms survives and becomes not small in the total energy.
In order to obtain the proper energy saturation variationally, it is necessary to include higher-body terms in the correlated Hamiltonian.

We investigated the advantage of the independent treatment of the correlation functions in TOAMD.
We perform constraint calculations in which the correlation functions are fixed to be common for every term of TOAMD.
This leads to an energy loss of few MeV for $s$-shell nuclei.
This result indicates that the correlation functions should be different at each order of the power series expansion in TOAMD.
From the variational point of view, this treatment is better than the case using common correlation function for all orders.
In fact, for the central MT-V potential, it is found that the numerical accuracy of TOAMD with double products of the correlation functions is beyond the variational Monte Carlo calculation 
using the product-type common correlation functions.

In future work, we shall increase the multiple products of the correlation functions to the triple case,
which is expected to increase the numerical accuracy of TOAMD.
The other way to extend TOAMD is the superposition of many AMD basis states.
Based on the success for the $s$-shell nuclei in the present analysis, we shall apply TOAMD to the $p$-shell nuclei.
We also have a plan to treat the three-nucleon interaction explicitly, e.g., the Fujita--Miyazawa type,
whose treatment is essentially similar to the case of the many-body operators of the correlated Hamiltonian in TOAMD.

\section*{Acknowledgements}
This work was supported by JSPS KAKENHI Grant Numbers JP15K05091, JP15K17662, and JP16K05351.
Numerical calculations were partially performed on a computer system at RCNP, Osaka University.

\appendix
\section{Matrix elements in TOAMD}\label{TOAMD_ME}

We explain briefly how to calculate the matrix elements of the operator $\hat O$ with the TOAMD wave function $\Phi_{\rm TOAMD}$;
these are equivalent to the matrix elements of the correlated operator $\tilde O$ with the AMD wave function $\Phi_{\rm AMD}$ as 
\begin{eqnarray}
    \bra \Phi_{\rm TOAMD}|\hat{O}   |\Phi_{\rm TOAMD} \ket
&=& \bra \Phi_{\rm AMD}  |\tilde {O}|\Phi_{\rm AMD}   \ket\,,
    \\
\tilde {O}&=& \hat O +F^\dagger \hat O + \hat O F + F^\dagger \hat O F +\cdots .
    \label{eq:correlated_O}
\end{eqnarray}
The correlated operator $\tilde O$ consists of the original operator $\hat O$, the multiple products of $\hat O$ and the two-body correlation function $F$.
Each term including $F$ in Eq.~(\ref{eq:correlated_O}) is expanded into a series of $n$-body operators in terms of the cluster expansion with $n=2,\cdots,A$.
We consider the case of the specific $n$-body operator $\tilde O^{[n]}$, 
which has the following form: $\tilde O^{[n]}=S^{[n]}\times \sum_{i_1,i_2,\ldots,i_n}^A \tilde O^{[n]}_{i_1 i_2 \ldots i_n}$ 
with symmetry factor $S^{[n]}$ for particle exchange and $n$-particle indices of $i_1,\cdots,i_n$.
The matrix elements of $\tilde O^{[n]}$ with the AMD wave function are given as
\begin{eqnarray}
    \bra \Phi_{\rm AMD} | \tilde O^{[n]} |\Phi_{\rm AMD} \ket
&=&	S^{[n]} \sum_{i_1,i_2,\ldots,i_n \atop j_1,j_2,\ldots,j_n }^A
	\bra \phi_{i_1}\phi_{i_2}\cdots \phi_{i_n} | \tilde O^{[n]}_{1\,2 \cdots n} | \phi_{j_1} \phi_{j_2} \cdots \phi_{j_n}  \ket
        \nonumber\\
&\times&{\rm det} \{B^{-1}_{j_1 i_1} B^{-1}_{j_2 i_2} \cdots B^{-1}_{j_ni_n}  \}\ {\rm det} B \,,
        \label{eq:kernel}
        \\
        B_{ij}
&=&     \bra \phi_{i}| \phi_{j} \ket,\qquad {\rm det} B~=~\bra \Phi_{\rm AMD} | \Phi_{\rm AMD} \ket \,,
        \label{eq:b_matrix}
\end{eqnarray}
where the quantity $B_{ij}$ is the single-nucleon overlap matrix element in AMD.

We shall consider the interaction $V$ as the operator $\hat{O}$. 
We express $V$ and $F$ in terms of the sum of Gaussian functions in coordinate space.
In order to calculate the kernel of the matrix elements
$\bra \phi_{i_1}\cdots\phi_{i_n} | \tilde O^{[n]}_{1\,2 \cdots n} | \phi_{j_1} \cdots \phi_{j_n}  \ket$ in Eq.~(\ref{eq:kernel}),
we use the Fourier transformation of the Gaussian for the interparticle coordinate $\vec r_{ij}=\vec r_i - \vec r_j$ with range $a$ 
in $V$ and $F$ \cite{myo15,goto79} as follows:
\begin{eqnarray}
   e^{-a\vec r_{ij}^{\,2}}
&=& \frac{1}{(2\pi)^3} \left(\frac{\pi}{a}\right)^{3/2} \int d \vec k\, e^{-\vec k^2/4a} \cdot 
    e^{i\vec k \cdot \vec r_i} \  e^{-i\vec k \cdot \vec r_j}\,,
   \label{eq:Fourier}
	\\
    {\vec r}_{ij} e^{-a \vec r_{ij}^{\,2}}
&=& \frac{1}{(2\pi)^3} \left(\frac{1}{2ai}\right) \left(\frac{\pi}{a}\right)^{3/2} \int d \vec k \, e^{-\vec k^2/4a} \cdot 
    e^{i\vec k \cdot \vec r_i} \  e^{-i\vec k \cdot \vec r_j} \cdot  \vec k \,,
   \label{eq:Fourier_r}
	\\
    {\vec r}_{ij}^{\,2} S_{12}(\hat r_{ij}) e^{-a \vec r_{ij}^{\,2}}
&=& \frac{1}{(2\pi)^3} \left(\frac{1}{2ai}\right)^2 \left(\frac{\pi}{a}\right)^{3/2} \int d \vec k \, e^{-\vec k^2/4a} \cdot 
    e^{i\vec k \cdot \vec r_i} \  e^{-i\vec k \cdot \vec r_j} \cdot
    \vec k^{2} S_{12}(\hat k) \,,
   \label{eq:Fourier_S12}
    \\
    S_{12}(\hat k)&=&3(\vec \sigma_i\cdot \hat k)(\vec \sigma_j \cdot \hat k)-\vec \sigma_i \cdot \vec \sigma_j~.
\end{eqnarray}
The tensor operator $S_{12}$ is also transformed into the momentum space. 
Equation~(\ref{eq:Fourier_r}) is used in the calculation of the $LS$ term.
The merit of this transformation is that the coordinate $\vec r_{ij}$ in the Gaussian function becomes separable for each particle coordinate $\vec r_i$ and $\vec r_j$ with the plane-wave form in momentum space.

We explain the case of the central interaction for $V$ and the central correlation for $F$ to calculate the kernel matrix elements. 
In the correlated interaction, $\tilde V$, the specific $n$-body operator $\tilde V^{[n]}$ becomes the products of several Gaussian functions and has various connections with the interparticle coordinates, 
such as the diagrams shown in Fig.~\ref{fig:FFF}.
We transform each Gaussian function in $\tilde V^{[n]}$ into the corresponding momentum space one by one with the Fourier transformation.
We denote the transformed operator in momentum space as $\tilde V^{[n]}_k$ which is a function of multi-momenta having Gaussian functions 
and includes the products of plane waves for each particle coordinate, as shown in Eq.~(\ref{eq:Fourier}).
The structure of $\tilde V^{[n]}_k$ for particle coordinates is schematically written as
\begin{eqnarray}
   \tilde V^{[n]}_k  
&=&  S^{[n]}\times \sum_{i_1,i_2,\ldots,i_n}^A \tilde V^{[n]}_{k,i_1 i_2 \ldots i_n} \,,
   \\
\tilde V^{[n]}_{k,1\, 2 \cdots n}   
&\propto&  g^{[n]}_1(\vec r_1) \cdot g^{[n]}_2(\vec r_2) \times \cdots \times g^{[n]}_n(\vec r_n) .   
\end{eqnarray}
Each operator $g^{[n]}_i(\vec r_i)$ has a form of $e^{i\vec K_i \cdot \vec r_i}$ depending on the momentum $\vec K_i$, which is the sum of the momenta for which the underlying Gaussian functions in Eq.~(\ref{eq:Fourier}) commonly involve the coordinate $\vec r_i$.
The operator $g^{[n]}_i(\vec r_i)$ can have spin and isospin operators, which are ignored at present.
Hence the kernel matrix elements of the $n$-body operator are given by the products of the single-particle matrix elements of the plane wave in AMD as
\begin{eqnarray}
     \bra \phi_{i_1}\cdots \phi_{i_n} | \tilde V^{[n]}_{k,1\, 2 \cdots n} | \phi_{j_1} \cdots \phi_{j_n}  \ket \
\propto
     \bra \phi_{i_1} | g^{[n]}_1(\vec r_1) |\phi_{j_1} \ket \times \cdots \times
     \bra \phi_{i_n} | g^{[n]}_n(\vec r_n) |\phi_{j_n} \ket ,
     \label{eq:ME_K}
\end{eqnarray}
where
\begin{eqnarray}
    \bra \phi_{i} | e^{i\vec K \cdot \vec r} | \phi_{j}  \ket
&=& \bra \phi_{i} | \phi_{j} \ket \cdot e^{ i\vec K \cdot (\vec D^*_i + \vec D_j) /2  - \vec K^2/8\nu} .
     \label{eq:ME_PW}
\end{eqnarray}
Finally we perform the multiple integration for all momenta analytically,
in which the integrand includes the Gaussian functions for each momentum.
For the kinetic energy and $LS$ force, we take the explicit derivatives by the corresponding coordinate \cite{myo15}.

We explicitly give the form of $g^{[n]}_i(\vec r_i)$ in the case of $F^\dagger F$ appearing in the correlated norm operator.
The operator $F^\dagger F$ gives two momenta of $\vec k_1$ and $\vec k_2$ from each $F$
and is expanded into three diagrams of $\frac12 [12:12]$, $[12:13]$, and $\frac14 [12:34]$ in the cluster expansion in Fig. \ref{fig:FF}.

For the configuration $\frac12 [12:12]$ as the two-body operator with ladder-type,
\begin{eqnarray}
g^{[2]}_1(\vec r_1)&=&e^{ i(\vec k_1+\vec k_2)\cdot \vec r_1},
\qquad
g^{[2]}_2(\vec r_2)~=~e^{-i(\vec k_1+\vec k_2)\cdot \vec r_2}.
\end{eqnarray}
For the configuration $[12:13]$ as the three-body operator linked via the coordinate $\vec r_1$,
\begin{eqnarray}
   g^{[3]}_1(\vec r_1)&=&e^{ i(\vec k_1+\vec k_2)\cdot \vec r_1},\qquad
   g^{[3]}_2(\vec r_2)~=~e^{-i \vec k_1\vec r_2},\qquad
   g^{[3]}_3(\vec r_3)~=~e^{-i \vec k_2\vec r_3}.
\end{eqnarray}
For the configuration $\frac14 [12:34]$ as the four-body operator with two unlinked operators:
\begin{eqnarray}
   g^{[4]}_1(\vec r_1)&=&e^{ i\vec k_1\cdot\vec r_1},\qquad
   g^{[4]}_2(\vec r_2)~=~e^{-i\vec k_1\cdot\vec r_2},
   \nonumber\\
   g^{[4]}_3(\vec r_3)&=&e^{ i\vec k_2\cdot\vec r_3},\qquad 
   g^{[4]}_4(\vec r_4)~=~e^{-i\vec k_2\cdot\vec r_4}.
\end{eqnarray}

\nc\PTEP[1]{Prog.\ Theor.\ Exp.\ Phys.,\ \andvol{#1}} 
\nc\PPNP[1]{Prog.\ Part.\ Nucl.\ Phys.,\ \andvol{#1}} 


\begin{thebibliography}{00}

\bibitem{pieper01}  S. C. Pieper and R. B. Wiringa, \JL{Annu. Rev. Nucl. Part. Sci.,51,53,2001}.
\bibitem{wiringa95} R. B. Wiringa, V.G.J. Stoks, and R. Schiavilla, \PRC{51,38,1995}
\bibitem{ikeda10}   K. Ikeda, T. Myo, K. Kat\=o, and H. Toki, in {\it Clusters in Nuclei} (Editors:Christian Beck, Springer, Berlin, 2010), Lecture Notes in Physics 818 Vol.1, pp.165-221.
\bibitem{ong13}     H. J. Ong et al., \PLB{725,277,2013}
\bibitem{myo05}     T. Myo, K. Kat\=o, and K. Ikeda, \PTP{113,763,2005}
\bibitem{myo07}     T. Myo, S. Sugimoto, K. Kat\=o, H. Toki, and K. Ikeda, \PTP{117,257,2007}
\bibitem{myo09}     T. Myo, H. Toki, and K. Ikeda, \PTP{121,511,2009}
\bibitem{feldmeier98} H. Feldmeier, T. Neff, R. Roth, and J. Schnack, \NPA{632,61,1998}
\bibitem{myo11}     T. Myo, A. Umeya, H. Toki, and K. Ikeda, \PRC{84,034315,2011}
\bibitem{myo12}     T. Myo, A. Umeya, H. Toki, and K. Ikeda, \PRC{86,024318,2012}
\bibitem{myo14}	    T. Myo, A. Umeya, K. Horii, H. Toki and K. Ikeda,~\PTEP{2014,033D01,2014}
\bibitem{myo15b}    T. Myo, A. Umeya, H. Toki, and K. Ikeda,~\PTEP{2015,063D03,2015}
\bibitem{ikeda68}     K. Ikeda, H. Horiuchi, and S. Saito, \PTPS{68,1,1980}
\bibitem{horiuchi12}  H. Horiuchi, K. Ikeda, and K. Kat\=o, \PTPS{192,1,2012}
\bibitem{barrett13}   B. R. Barrett, P. Navr\'atil, and J. P. Vary, \PPNP{69,131,2013}
\bibitem{myo15}       T. Myo, H. Toki, K. Ikeda, H. Horiuchi, and T. Suhara, \PTEP{2015,073D02,2015}
\bibitem{myo16}       T. Myo, H. Toki, K. Ikeda, H. Horiuchi, and T. Suhara, \PLB{769,213,2017}
\bibitem{myo17}       T. Myo, H. Toki, K. Ikeda, H. Horiuchi, and T. Suhara, \PRC{95,044314,2017}
\bibitem{kanada03}    Y. Kanada-En'yo, M. Kimura, and H. Horiuchi, \JL{C. R. Phys.,4,497,2003}.
\bibitem{kanada12}    Y. Kanada-En'yo, M. Kimura, and A. Ono, \PTEP{2012,01A202,2012}
\bibitem{toki17}      H. Toki and J. Hu, \JL{Chin. J. Phys.,55,28,2017}.
\bibitem{kamada01}    H. Kamada et al.,~\PRC{64,044001,2001} and references therein.
\bibitem{malfliet69}  R. A. Malfliet and J. A. Tjon, \NPA{127,161,1969}
\bibitem{sugie57}     A. Sugie, P. E. Hodgson, and H. H. Robertson, \JL{Proc. Phys. Soc.,70A,1,1957}.
\bibitem{nagata59}    S. Nagata, T. Sasakawa, T. Sawada, and R.Tamagaki, \PTP{22,274,1959}
\bibitem{bishop92}    R. F. Bishop, E. Buend\'ia, M. F. Flynnsl, and R. Guardiola, \JPG{18,1157,1992}
\bibitem{bishop98}    R. F. Bishop, E. Buend\'ia, M. F. Flynnsl, and R. Guardiola, \NPA{643,243,1998}
\bibitem{goto79}      Y. Goto and H. Horiuchi, \PTP{62,662,1979}
\bibitem{zabolitzky82}J. Zabolitzky, K. E. Schmidt, and M. H. Kalos, \PRC{25,1111,1982}
\bibitem{carlson81}   J. Carlson and V. R. Pandharipande, \NPA{371,301,1981}
\bibitem{varga95}     K. Varga and Y. Suzuki, \PRC{52,2885,1995} and references therein.
\bibitem{fukukawa15}  K. Fukukawa, M. Baldo, G. F. Burgio, L. Lo Monaco, and H.-J. Schulze, \PRC{92,065802,2015}
\bibitem{suzuki08}    Y. Suzuki, W. Horiuchi, M. Orabi, and K. Arai, \JL{Few-Body Systems, 42,33,2008}.
\end{thebibliography}
\end{document}